\documentclass[preprint2]{aastex63}
\usepackage{float, bm, graphicx, amsmath, morefloats, enumitem}
\usepackage{lineno}

\submitjournal{ApJ}

\shorttitle{SN\,2020bvc}
\shortauthors{Ho et al.}

\newcommand{\nickel}{\mbox{$\rm {}^{56}Ni$}}
\newcommand{\cobalt}{\mbox{$\rm {}^{56}Co$}}

\newcommand{\chandra}{{\em Chandra}}
\newcommand{\swift}{{\em Swift}}
\newcommand{\fermi}{{\em Fermi}}

\newcommand{\kelvin}{\mbox{$\rm K$}}

\newcommand{\erg}{\mbox{$\rm erg$}}
\newcommand{\kev}{\mbox{$\rm keV$}}
\newcommand{\jy}{\mbox{$\rm Jy$}}

\newcommand{\gauss}{\mbox{$\rm G$}}

\newcommand{\cm}{\mbox{$\rm cm$}}
\newcommand{\km}{\mbox{$\rm km$}}

\newcommand{\pcmsq}{\mbox{$\rm cm^{-2}$}}
\newcommand{\cmsq}{\mbox{$\rm cm^{2}$}}
\newcommand{\degsq}{\mbox{$\rm deg^{2}$}}

\newcommand{\pcmcub}{\mbox{$\rm cm^{-3}$}}

\newcommand{\pgpccub}{\mbox{$\rm Gpc^{-3}$}}

\newcommand{\msol}{\mbox{$\rm M_\odot$}}
\newcommand{\mni}{\mbox{$\rm M_{Ni}$}}
\newcommand{\mej}{\mbox{$\rm M_{ej}$}}
\newcommand{\pgram}{\mbox{$\rm g^{-1}$}}

\newcommand{\days}{\mbox{$\rm d$}}
\newcommand{\psec}{\mbox{$\rm s^{-1}$}}
\newcommand{\pyr}{\mbox{$\rm yr^{-1}$}}

\newcommand{\ghz}{\mbox{$\rm GHz$}}
\newcommand{\hz}{\mbox{$\rm GHz$}}
\newcommand{\phz}{\mbox{$\rm Hz^{-1}$}}

\begin{document}

\title{
SN\,2020bvc: a Broad-lined Type Ic Supernova with a Double-peaked Optical Light Curve and a Luminous X-ray and Radio Counterpart}

\author[0000-0002-9017-3567]{Anna Y. Q.~Ho}
\affiliation{Cahill Center for Astrophysics, 
California Institute of Technology, MC 249-17, 
1200 E California Boulevard, Pasadena, CA, 91125, USA}

\author[0000-0001-5390-8563]{S. R.~Kulkarni}
\affiliation{Cahill Center for Astrophysics, 
California Institute of Technology, MC 249-17, 
1200 E California Boulevard, Pasadena, CA, 91125, USA}

\author[0000-0001-8472-1996]{Daniel A.~Perley}
\affiliation{Astrophysics Research Institute, Liverpool John Moores University, IC2, Liverpool Science Park, 146 Brownlow Hill, Liverpool L3 5RF, UK}

\author[0000-0003-1673-970X]{S.\ Bradley~Cenko}
\affiliation{Astrophysics Science Division, NASA Goddard Space Flight Center, Mail Code 661, Greenbelt, MD 20771, USA}
\affiliation{Joint Space-Science Institute, University of Maryland, College Park, MD 20742, USA}

\author{Alessandra Corsi}
\affiliation{Department of Physics and Astronomy, Texas Tech University, Box 1051, Lubbock, TX 79409-1051, USA}

\author[0000-0001-6797-1889]{Steve Schulze}
\affiliation{Department of Particle Physics and Astrophysics, Weizmann Institute of Science, 234 Herzl St, 76100 Rehovot, Israel}

\author[0000-0001-9454-4639]{Ragnhild Lunnan}
\affiliation{The Oskar Klein Centre \& Department of Astronomy, Stockholm University, AlbaNova, SE-106 91 Stockholm, Sweden}

\author[0000-0003-1546-6615]{Jesper Sollerman}
\affiliation{The Oskar Klein Centre \& Department of Astronomy, Stockholm University, AlbaNova, SE-106 91 Stockholm, Sweden}

\author{Avishay Gal-Yam}
\affiliation{Department of Particle Physics and Astrophysics, Weizmann Institute of Science, 234 Herzl St, 76100 Rehovot, Israel}

\author{Shreya Anand}
\affiliation{Division of Physics, Mathematics, and Astronomy, California Institute of Technology, Pasadena, CA 91125, USA}

\author[0000-0002-3821-6144]{Cristina Barbarino}
\affiliation{The Oskar Klein Centre \& Department of Astronomy, Stockholm University, AlbaNova, SE-106 91 Stockholm, Sweden}

\author[0000-0001-8018-5348]{Eric C. Bellm}
\affiliation{DIRAC Institute, Department of Astronomy, University of Washington, 3910 15th Avenue NE, Seattle, WA 98195, USA}

\author[0000-0001-8208-2473]{Rachel J. Bruch}
\affiliation{Department of Particle Physics and Astrophysics, Weizmann Institute of Science, Rehovot 76100, Israel}

\author{Eric Burns}
\affiliation{NASA Postdoctoral Program Fellow, Goddard Space Flight Center, Greenbelt, MD 20771, USA}
\affiliation{Department of Physics \& Astronomy, Louisiana State University, Baton Rouge, LA 70803, USA}

\author{Kishalay De}
\affiliation{Cahill Center for Astrophysics, 
California Institute of Technology, MC 249-17, 
1200 E California Boulevard, Pasadena, CA, 91125, USA}

\author[0000-0002-5884-7867]{Richard Dekany}
\affiliation{Caltech Optical Observatories, California Institute of Technology, Pasadena, CA  91125}

\author{Alexandre Delacroix}
\affiliation{Caltech Optical Observatories, California Institute of Technology, Pasadena, CA  91125}

\author[0000-0001-5060-8733]{Dmitry A. Duev}
\affiliation{Cahill Center for Astrophysics, 
California Institute of Technology, MC 249-17, 
1200 E California Boulevard, Pasadena, CA, 91125, USA}

\author[0000-0002-1153-6340]{Dmitry D. Frederiks}
\affiliation{Ioffe Institute, Politekhnicheskaya 26, St. Petersburg 194021, Russia}

\author[0000-0002-4223-103X]{Christoffer Fremling}
\affiliation{Cahill Center for Astrophysics, 
California Institute of Technology, MC 249-17, 
1200 E California Boulevard, Pasadena, CA, 91125, USA}

\author[0000-0003-3461-8661]{Daniel A.~Goldstein}
\altaffiliation{Hubble Fellow}
\affiliation{Cahill Center for Astrophysics, 
California Institute of Technology, MC 249-17, 
1200 E California Boulevard, Pasadena, CA, 91125, USA}

\author[0000-0001-8205-2506]{V. Zach Golkhou}
\affiliation{DIRAC Institute, Department of Astronomy, University of Washington, 3910 15th Avenue NE, Seattle, WA 98195, USA} 
\affiliation{The eScience Institute, University of Washington, Seattle, WA 98195, USA}
\altaffiliation{Moore-Sloan, WRF Innovation in Data Science, and DIRAC Fellow}

\author{Matthew J. Graham}
\affiliation{Cahill Center for Astrophysics, 
California Institute of Technology, MC 249-17, 
1200 E California Boulevard, Pasadena, CA, 91125, USA}

\author{David Hale}
\affiliation{Caltech Optical Observatories, California Institute of Technology, Pasadena, CA  91125}

\author{Mansi M.~Kasliwal}
\affiliation{Cahill Center for Astrophysics, 
California Institute of Technology, MC 249-17, 
1200 E California Boulevard, Pasadena, CA, 91125, USA}

\author[0000-0002-6540-1484]{Thomas Kupfer}
\affiliation{Kavli Institute for Theoretical Physics, University of California, Santa Barbara, CA 93106, USA}

\author[0000-0003-2451-5482]{Russ R. Laher}
\affiliation{IPAC, California Institute of Technology, 1200 E. California
             Blvd, Pasadena, CA 91125, USA}
             
\author[0000-0003-2211-4001]{Julia Martikainen}
\affiliation{Department of Physics, University of Helsinki, P.O. box 64, FI-00014, Finland}
\affiliation{Nordic Optical Telescope,  Roque de Los Muchachos Observatory, Rambla José Ana Fernández Pérez 7, 38711 Breña Baja, La Palma, Canarias, Spain}

\author[0000-0002-8532-9395]{Frank J. Masci}
\affiliation{IPAC, California Institute of Technology, 1200 E. California
             Blvd, Pasadena, CA 91125, USA}
             
\author{James D. Neill}
\affiliation{Cahill Center for Astrophysics, 
California Institute of Technology, MC 249-17, 
1200 E California Boulevard, Pasadena, CA, 91125, USA}

\author{Anna Ridnaia}
\affiliation{Ioffe Institute, Politekhnicheskaya 26, St. Petersburg 194021, Russia}

\author[0000-0001-7648-4142]{Ben Rusholme}
\affiliation{IPAC, California Institute of Technology, 1200 E. California
             Blvd, Pasadena, CA 91125, USA}

\author{Volodymyr Savchenko}
\affiliation{ISDC, Department of Astronomy, University of Geneva, Chemin d'Ecogia, 16 CH-1290 Versoix, Switzerland}

\author[0000-0003-4401-0430]{David L. Shupe}
\affiliation{IPAC, California Institute of Technology, 1200 E. California
             Blvd, Pasadena, CA 91125, USA}
             
\author[0000-0001-6753-1488]{Maayane T. Soumagnac}
\affiliation{Lawrence Berkeley National Laboratory, 1 Cyclotron Road, Berkeley, CA 94720, USA}
\affiliation{Department of Particle Physics and Astrophysics, Weizmann Institute of Science, Rehovot 76100, Israel}

\author{Nora L.~Strotjohann}
\affiliation{Department of Particle Physics and Astrophysics, Weizmann Institute of Science, 234 Herzl St, 76100 Rehovot, Israel}

\author[0000-0002-2208-2196]{Dmitry S. Svinkin}
\affiliation{Ioffe Institute, Politekhnicheskaya 26, St. Petersburg 194021, Russia}

\author{Kirsty Taggart}
\affiliation{Astrophysics Research Institute, Liverpool John Moores University, IC2, Liverpool Science Park, 146 Brownlow Hill, Liverpool L3 5RF, UK}

\author{Leonardo Tartaglia}
\affiliation{The Oskar Klein Centre \& Department of Astronomy, Stockholm University, AlbaNova, SE-106 91 Stockholm, Sweden}

\author[0000-0003-1710-9339]{Lin Yan}
\affiliation{Caltech Optical Observatories, California Institute of Technology, Pasadena, CA  91125}

\author{Jeffry Zolkower}
\affiliation{Caltech Optical Observatories, California Institute of Technology, Pasadena, CA  91125}

\begin{abstract}

We present optical, radio, and X-ray observations of SN\,2020bvc (=ASASSN20bs; ZTF20aalxlis),
a nearby ($z=0.0252$; $d=114$\,Mpc) broad-lined (BL) Type Ic supernova (SN) and the first double-peaked Ic-BL  discovered without a gamma-ray burst (GRB) trigger.
Our observations show that SN\,2020bvc shares several properties in common with the Ic-BL SN\,2006aj, which was associated with the low-luminosity gamma-ray burst (LLGRB)\,060218.
First, the 10\,GHz radio luminosity ($L_\mathrm{radio} \approx 10^{37}\,\erg\,\psec$) is brighter than ordinary core-collapse SNe but fainter than LLGRB-SNe such as SN\,1998bw (associated with LLGRB\,980425).
We model our VLA observations (spanning 13--43\,d) as synchrotron emission from a mildly relativistic ($v \gtrsim 0.3c$) forward shock.
Second, with \swift\ and \chandra\ we detect X-ray emission ($L_X \approx 10^{41}\,\erg\,\psec$) that is not naturally explained as inverse Compton emission or as part of the same synchrotron spectrum as the radio emission.
Third, high-cadence ($6\times$/night) data from the Zwicky Transient Facility (ZTF) shows a double-peaked optical light curve, the first peak from shock-cooling of extended low-mass material (mass $M_e<10^{-2}\,\msol$ at radius $R_e>10^{12}\,\cm$) and the second peak from the radioactive decay of \nickel.
SN\,2020bvc is the first double-peaked Ic-BL SN discovered without a GRB trigger, so it is noteworthy that it shows X-ray and radio emission similar to LLGRB-SNe.
For four of the five other nearby ($z\lesssim0.05$) Ic-BL SNe with ZTF high-cadence data, we rule out a first peak like that seen in SN\,2006aj and SN\,2020bvc, i.e. that lasts
$\approx 1\,\days$ and reaches a peak luminosity $M \approx -18$.
X-ray and radio follow-up observations of Ic-BL SNe with well-sampled early optical light curves will establish whether double-peaked optical light curves are indeed predictive of LLGRB-like X-ray and radio emission.

\end{abstract}


\section{Introduction}
\label{sec:intro}

It is well-established that most
long-duration gamma-ray bursts (GRBs) arise from massive-star explosions
(see \citet{Woosley2006} for a detailed review,
and \citet{Hjorth2012} and \citet{Cano2017} for recent updates).
The traditional model (reviewed in \citealt{Piran2004}) is that a massive star, stripped of its hydrogen and helium envelopes, collapses and forms a black hole or neutron star.
Through rotational spindown or accretion, the newborn compact object launches an outflow that tunnels through the star, breaks out from the surface as a narrowly collimated jet, and appears as a GRB when viewed on-axis from Earth.
The jet shocks the circumburst medium,
producing a long-lived ``afterglow'' across the electromagnetic spectrum.
The same ``central engine'' that launches the GRB also
unbinds the stellar material in a supernova (SN) that has a greater kinetic energy ($10^{52}\,$erg) and photospheric velocity ($\gtrsim 20,000\,\km\,\psec$) than ordinary core-collapse SNe do \citep{Sobacchi2017,Barnes2018}.
These high-velocity, high-energy SNe were originally called
``hypernovae'' (e.g. \citealt{Iwamoto1998})
but a more common term today is ``broad-lined Type Ic'' (Ic-BL) SNe \citep{GalYam2017}.

Thousands of GRBs have been discovered,
with hundreds of afterglows and a dozen Ic-BL SNe (GRB-SNe) identified in follow-up observations.
Half of known GRB-SNe 
are associated with \emph{low-luminosity} GRBs (LLGRBs), defined as having an isotropic gamma-ray luminosity of $L_{\gamma,\mathrm{iso}}<10^{48.5}\,\erg\,\psec$
rather than the 
$L_{\gamma,\mathrm{iso}}>10^{49.5}\,\erg\,\psec$
of cosmological GRBs \citep{Hjorth2013,Cano2017}.
Although LLGRBs are 10--100 times more common than cosmological GRBs \citep{Soderberg2006aj,Liang2007},
the discovery rate by GRB detectors is much lower (one every few years)
due to the small volume in which they can be detected.
So, the sample size remains small,
and the connection between classical GRBs, LLGRBs, and Ic-BL SNe remains unknown.

To make progress on understanding the GRB-LLGRB-SN connection, wide-field high-cadence optical surveys can be used in conjunction with radio and X-ray follow-up observations to discover GRB-related phenomena without relying on a GRB trigger (e.g. \citealt{Soderberg2010,Cenko2013,Margutti2014,Corsi2017}).
To this end, for the past two years we have been conducting a systematic search for engine-driven explosions using the high-cadence
($6\times$/night) and nightly cadence ($2\times$/night) surveys of
the Zwicky Transient Facility (ZTF; \citealt{Graham2019,Bellm2019}),
 which have a combined area of 5000\,\degsq\ \citep{Bellm2019-sched}.

Here we present the most recent event detected as part of the ZTF engine-driven SN program: SN\,2020bvc (=ASASSN-20bs) was
first reported to the Transient Name Server (TNS\footnote{\url{https://wis-tns.weizmann.ac.il/}}) by the All-Sky Automated Survey for SuperNovae (ASAS-SN; \citealt{Shappee2014}), and
the discovery announcement noted the rapid rise and likely core-collapse (CC) SN origin \citep{Stanek2020}.
It was also reported by the Asteroid Terrestrial-impact Last Alert System   
(ATLAS; \citealt{Tonry2018}) as ATLAS20feh (on Feb 05.61).
The first detection of SN\,2020bvc was in
ZTF high-cadence data on Feb 04.34.
We classified the event as a Type Ic-BL SN \citep{Perley2020},
and the high-cadence data showed a double-peaked light curve.
Recognizing the similarity to
the Ic-BL SN\,2006aj associated with LLGRB\,060218 \citep{Soderberg2006aj,Mirabal2006,Pian2006,Sollerman2006,Ferrero2006},
we triggered X-ray \citep{Ho2020-chandra} and radio \citep{Ho2020-vla} follow-up observations.

This paper is structured as follows.
We present our observations of SN\,2020bvc in \S\ref{sec:obs}.
In \S\ref{sec:lc-analysis} we
measure basic light-curve properties and the blackbody evolution.
In \S\ref{sec:analysis-spec} we discuss the evolution of the optical spectra.
In \S\ref{sec:modeling-opt-lc} we show that the optical light curve can be explained as a combination of shock-cooling emission from extended low-mass material ($M_e<10^{-2}\,\msol$ at $R_e>10^{12}\,\cm$) and radioactive decay of \nickel.
In \S\ref{sec:modeling-radio} we model the forward shock,
and show that the radio emission can be explained with
velocities that are only mildly relativistic.
In \S\ref{sec:early-ztf-lc} we show ZTF light curves of five other nearby ($z<0.05$) Ic-BL SNe in the high-cadence surveys, and rule out a luminous first peak like that seen in SN\,2006aj and SN\,2020bvc for four events.
We conclude in \S\ref{sec:summary} by summarizing the properties of SN\,2020bvc and discussing its implications for the GRB-LLGRB-SN connection.

\section{Observations}
\label{sec:obs}

\subsection{ZTF Detection and Classification}
\label{sec:ztf-detection}

SN\,2020bvc was first detected
on 2020 Feb 04.34\footnote{All times given in UTC}
at $i=17.48\pm0.05\,$mag\footnote{All magnitudes given in AB}
at $\alpha = 14^{\mathrm{h}}33^{\mathrm{m}}57\fs01$, $\delta = +40^{\mathrm{d}}14^{\mathrm{m}}37\fs5$ (J2000),
as part of the ZTF Uniform Depth Survey\footnote{45 fields (2000\,\degsq) twice per night in each of $g$-, $r$-, and $i$-band} (Goldstein et al. in prep) with the 48-inch Samuel Oschin Schmidt telescope at Palomar Observatory (P48).
The ZTF observing system is described in \citet{Dekany2020}.
The identification of SN\,2020bvc made use of machine learning-based real-bogus classifiers \citep{Mahabal2019,Duev2019} and a star-galaxy separator \citep{Tachibana2018}.

The last non-detection by ZTF was 1.78\,d prior ($r>20.67\,$mag),
with more recent limits from 
ATLAS (0.67\,\days, $o>19.4\,$mag) and ASAS-SN (0.74\,\days, $g>18.6\,$mag).
Throughout the paper, we use the time of the ATLAS non-detection (Feb 03.67) as our reference epoch $t_0$.
Our estimate of the ``epoch of first light'' $t_0$ is
supported by aligning the light curves of SN\,2020bvc and SN\,2006aj, 
discussed in \S\ref{sec:lc-comparisons}.

Two hours after the first detection,
we obtained a spectrum using
the Spectral Energy Distribution Machine (SEDM; \citealt{Blagorodnova2018,Rigault2019}),
a low-resolution spectrograph on the automated 60-inch telescope at Palomar Observatory (P60; \citealt{Cenko2006}).
The spectrum is dominated by a thermal continuum,
with hydrogen emission lines from the host galaxy and possible weak absorption features that we discuss in \S\ref{sec:analysis-spec}.
On Feb 08.24, a spectrum we obtained using the
Spectrograph for the Rapid Acquisition of Transients (SPRAT; \citealt{Piascik2014}) on the Liverpool Telescope (LT; \citealt{Steele2004}) showed features consistent with a Type Ic-BL SN \citep{Perley2020}.
We discuss the spectroscopic evolution of SN\,2020bvc in \S\ref{sec:analysis-spec}. Follow-up observations were coordinated through the GROWTH Marshal \citep{Kasliwal2019}, and
the optical photometry and spectroscopy will be made public through WISeREP, the Weizmann Interactive Supernova Data Repository \citep{Yaron2012}.

\subsection{Host Galaxy}
\label{sec:host}

The position of SN\,2020bvc is 13\arcsec\ (7.2\,kpc\footnote{$\Lambda$CDM cosmology of \citet{Planck2016} used throughout.}) from the center of UGC\,09379
($z=0.025201\pm 0.000021$ from NED\footnote{\url{ned.ipac.caltech.edu}}),
which also hosted PTF13ast \citep{GalYam2014}.
UGC\,09379 is a massive galaxy: \cite{Chang2015} estimate a stellar mass $\log_{10}(M/\msol) = 10.28^{+0.01}_{-0.16}$ while the NASA-SDSS Atlas value (\citealt{Blanton2011}) is $\log_{10}(M/\msol) = 10.26$, comparable to the Milky Way and other large spirals. The stellar mass of UGC\,09379
is larger than that of all known GRB-SN host galaxies \citep{Hjorth2012,Taggart2019}
and similar only to the host galaxy of LLGRB\,171205A/SN\,2017iuk \citep{Delia2018,Wang2018,Izzo2019},
which had $\log_{10}(M/\msol) = 10.1\pm0.1$ \citep{Perley2017-host}.
The stellar mass of UGC\,09379 is also
larger than that of most Ic-BL SN host galaxies \citep{Modjaz2020},
with the exception of SN\,2002ap\footnote{M74: $\log_{10}(M/\msol) = 11.52^{+0.05}_{-0.05}$ \citep{Kelly2012}} and SN\,1997ef\footnote{
UGC\,4107:
$\log_{10}(M/\msol) = 10.55^{+0.07}_{-0.56}$ \citep{Kelly2012}}.



\begin{figure}[htb!]
    \centering
    \includegraphics[width=0.8\linewidth]{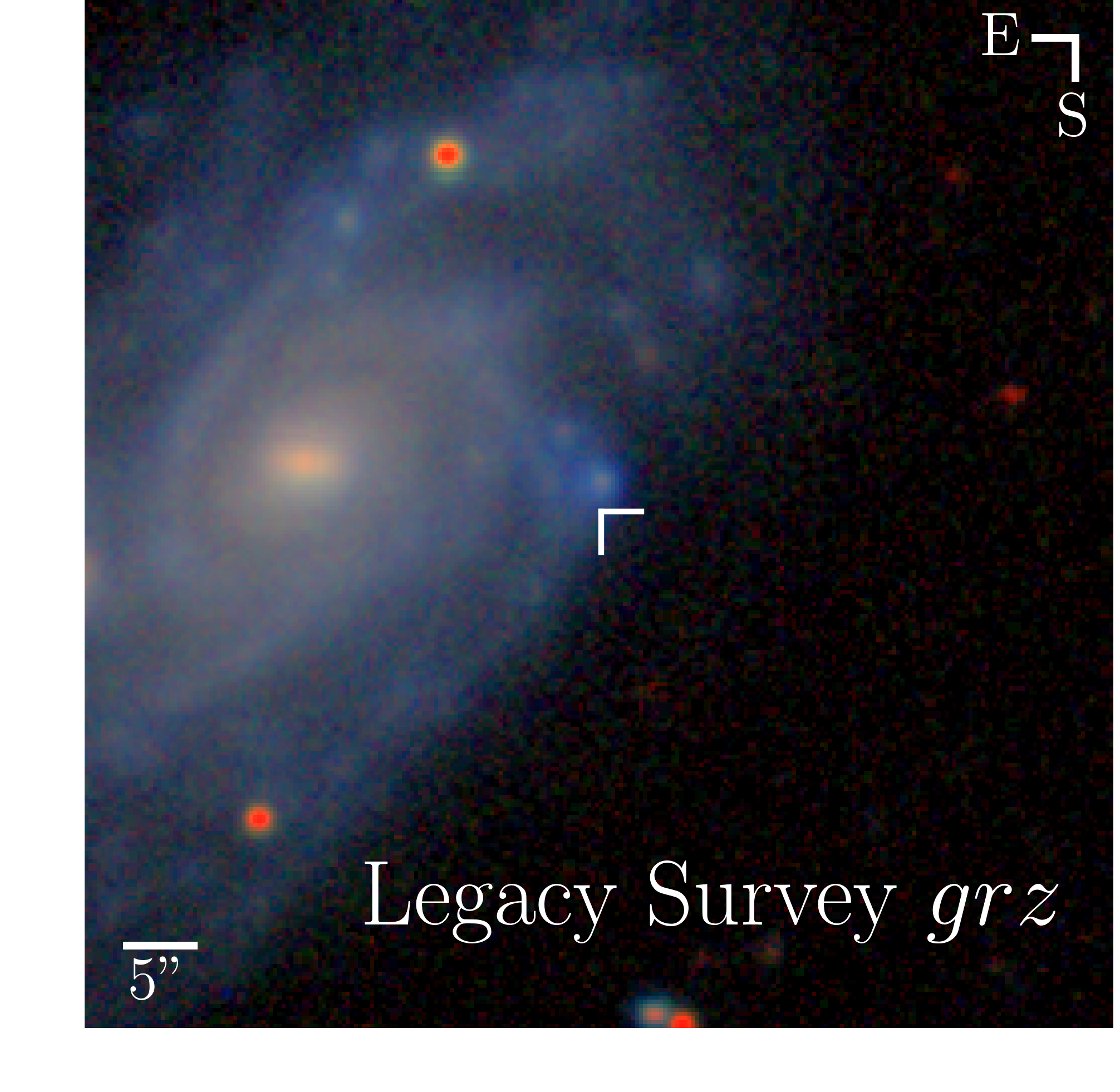}
    \caption{The position of SN\,2020bvc (white crosshairs) in its host galaxy UGC\,09379. $g$-, $r$-, and $z$-band images from the DESI Legacy Survey \citep{Dey2019} were combined using
    the prescription in \citet{Lupton2004}.}
    \label{fig:host}
\end{figure}

As shown in Figure~\ref{fig:host},
SN\,2020bvc is $1\farcs46\pm0\farcs34$ ($804 \pm 187\,$pc) from a bright \ion{H}{2} region.
We leave a detailed analysis of the SN site to future work,
but note that two nearby LLGRB-SNe,
LLGRB\,980425/SN\,1998bw \citep{Galama1998,Kulkarni1998} and LLGRB\,020903 \citep{Soderberg2004,Sakamoto2004,Bersier2006},
were located 800\,pc and 460\,pc, respectively, from similar bright compact regions in the outskirts of their host galaxies \citep{Sollerman2005,Hammer2006}.
Because these events took place outside the nearest massive-star cluster, it has been argued that the progenitors were Wolf-Rayet stars ejected from the cluster \citep{Hammer2006,Cantiello2007,Eldridge2011,vandenHeuvel2013}.  We also note that the metallicity of the SN site is quite low, making the appearance of a SN of this type in such a massive galaxy less surprising. We infer 12+log[O/H] = 8.2 from the underlying nebular emission in our March 22nd LRIS spectrum using the N2 diagnostic from \citealt{PP04},
consistent with the measurement of \citet{Izzo2020}.

\subsection{Optical Photometry}
\label{sec:phot}

\begin{figure*}[htb!]
    \centering
    \includegraphics[width=\textwidth]{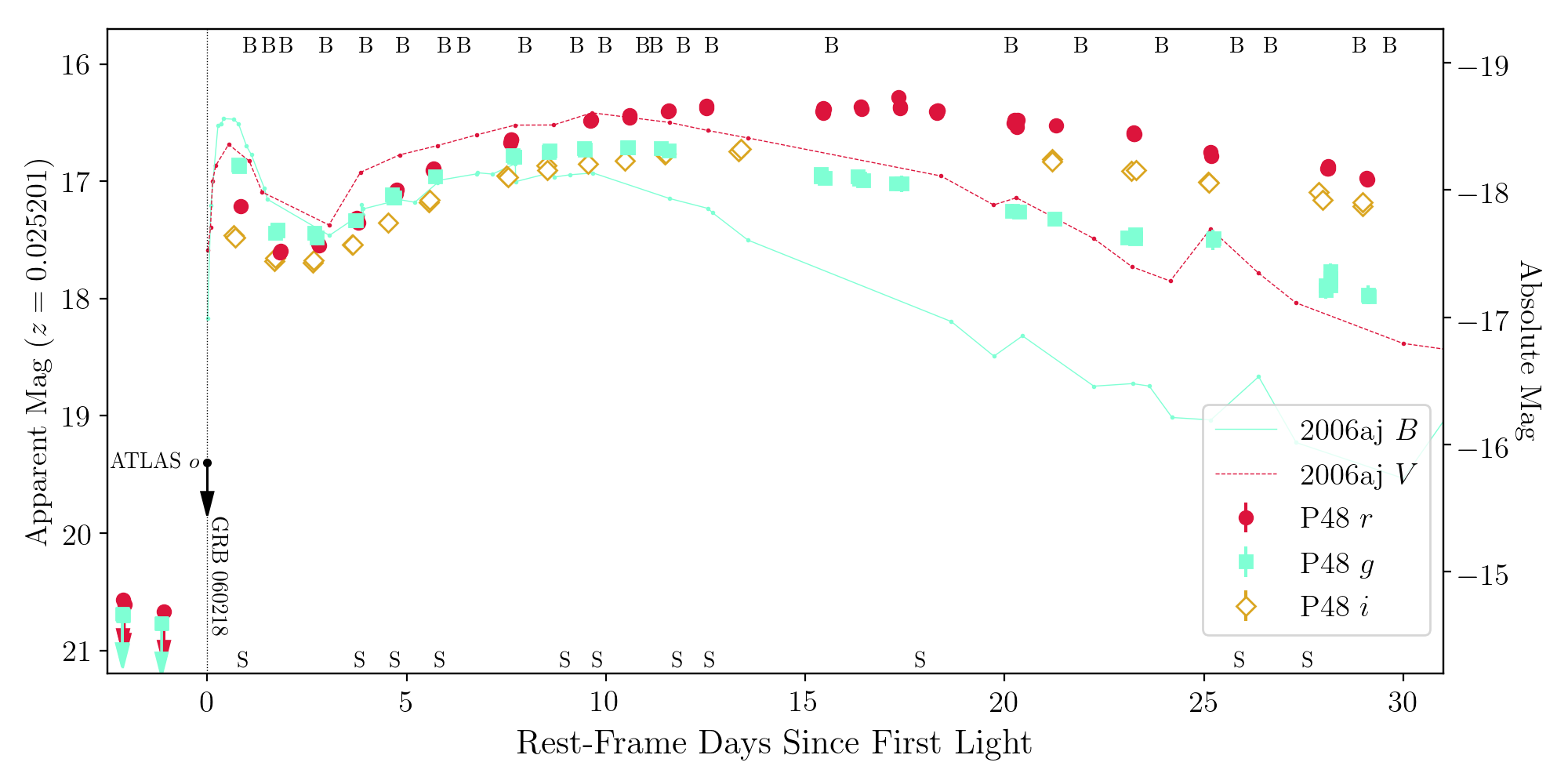}
    \caption{$g$-, $r$-, and $i$-band light curves of SN\,2020bvc from the ZTF Uniform Depth Survey (ZUDS),
    and an upper limit from ATLAS.
    Measurements have been corrected for Milky Way extinction.
    Epochs of follow-up spectroscopy are indicated with an `S' along the bottom of the figure.
    Epochs of blackbody fits (\S\ref{sec:bbfits}) are indicated with `B' along the top of the figure.
    For comparison, we show $B$ and $V$-band light curves of SN\,2006aj ($z=0.033$) transformed to the redshift of SN\,2020bvc ($z=0.025201$).
    The SN\,2006aj light curve was taken from the Open Supernova Catalog and corrected for Milky Way extinction; the data is originally from \citet{Modjaz2006}, \citet{Bianco2014}, and \citet{Brown2014}.
    We indicate the relative time of LLGRB\,060218 compared to the light curve of SN\,2006aj.}
    \label{fig:full-lc}
\end{figure*}

\begin{figure*}
    \centering
    \includegraphics[width=\textwidth]
    {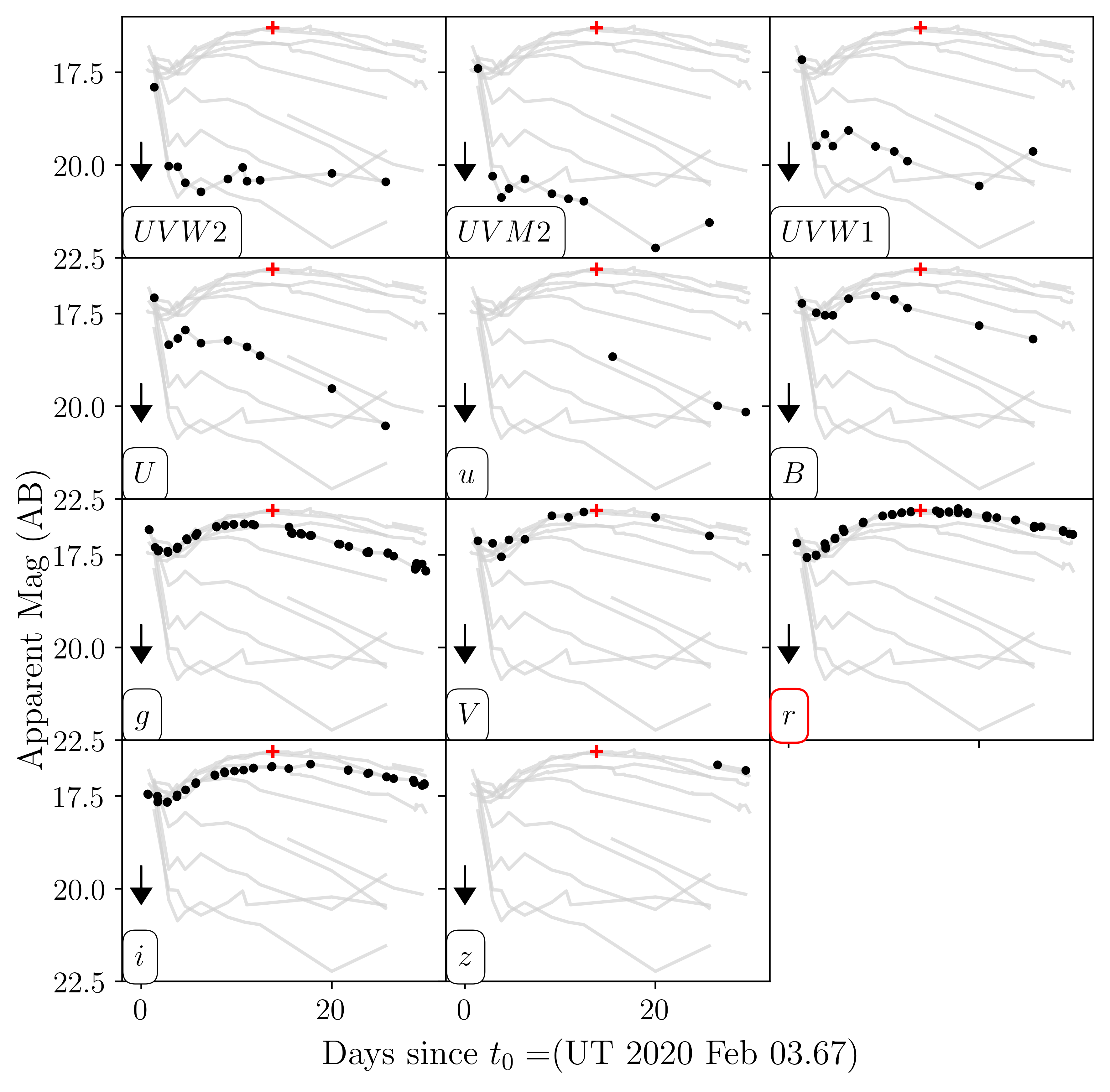}
    \caption{UV and optical light curves of SN\,2020bvc from \emph{Swift} and ground-based facilities. The arrow marks the last upper limit, which was by ATLAS in $o$-band. The red cross marks the peak of the $r$-band light curve. The full set of lightcurves is shown as grey lines in the background, and each panel highlights an individual filter in black.}
    \label{fig:lc-grid}
\end{figure*}

As shown in Figure~\ref{fig:full-lc},
SN\,2020bvc was observed almost nightly in $gri$ by the P48 for the first month post-explosion.
We obtained additional $ugriz$ and $gri$ photometry using the IO:O on LT and the SEDM on the P60, respectively.
The pipeline for P48 photometry is described in
\citet{Masci2019}, and makes use of the the image subtraction method of \citet{Zackay2016}.
LT image reduction was provided by the basic IO:O pipeline. P60 and LT image subtraction
were performed following \citet{Fremling2016},
using PS1 images for $griz$ and SDSS for $u$-band.
Values were corrected for Milky Way extinction
following \citet{Schlafly2011} with $E(B-V)=A_V/R_V=0.034\,$mag,
using $R_V=3.1$ and a \citet{Fitzpatrick1999} extinction law.
The full set of photometry is provided in Table~\ref{tab:uvot-phot} in Appendix~\ref{sec:appendix-phot-table},
and plotted in Figure~\ref{fig:lc-grid}.

\subsection{Spectroscopy}
\label{sec:spec}

We obtained 13 ground-based optical spectra
using the SEDM,
the Andalusia Faint Object Spectrograph and Camera (ALFOSC\footnote{http://www.not.iac.es/instruments/alfosc/}) on the Nordic Optical Telescope (NOT; \citealt{Djupvik2010}),
the Double Beam Spectrograph (DBSP; \citealt{Oke1982}) on the 200-inch Hale telescope at Palomar Observatory,
SPRAT on LT,
and the Low Resolution Imaging Spectrometer (LRIS; \citealt{Oke1995}) on the Keck~I 10-m telescope.
The SEDM pipeline is described in \citet{Rigault2019}, the SPRAT pipeline is based on the FrodoSpec pipeline \citep{Barnsley2012}, the P200/DBSP pipeline is described in \citet{Bellm2016}, and the Keck/LRIS pipeline \texttt{Lpipe} is described in \citet{Perley2019lpipe}.

\begin{figure*}[htb!]
\gridline{\fig{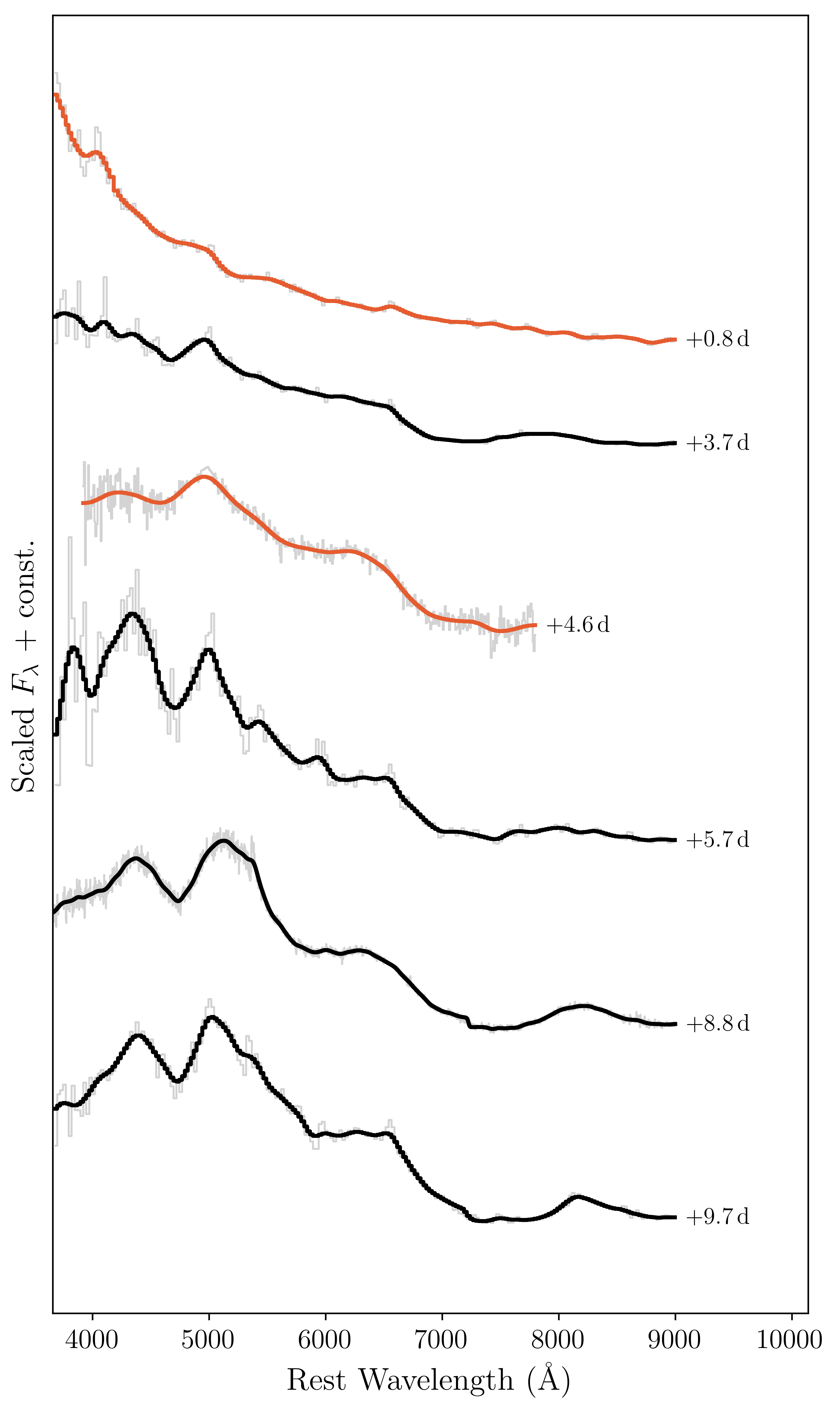}{0.47\textwidth}{}
          \fig{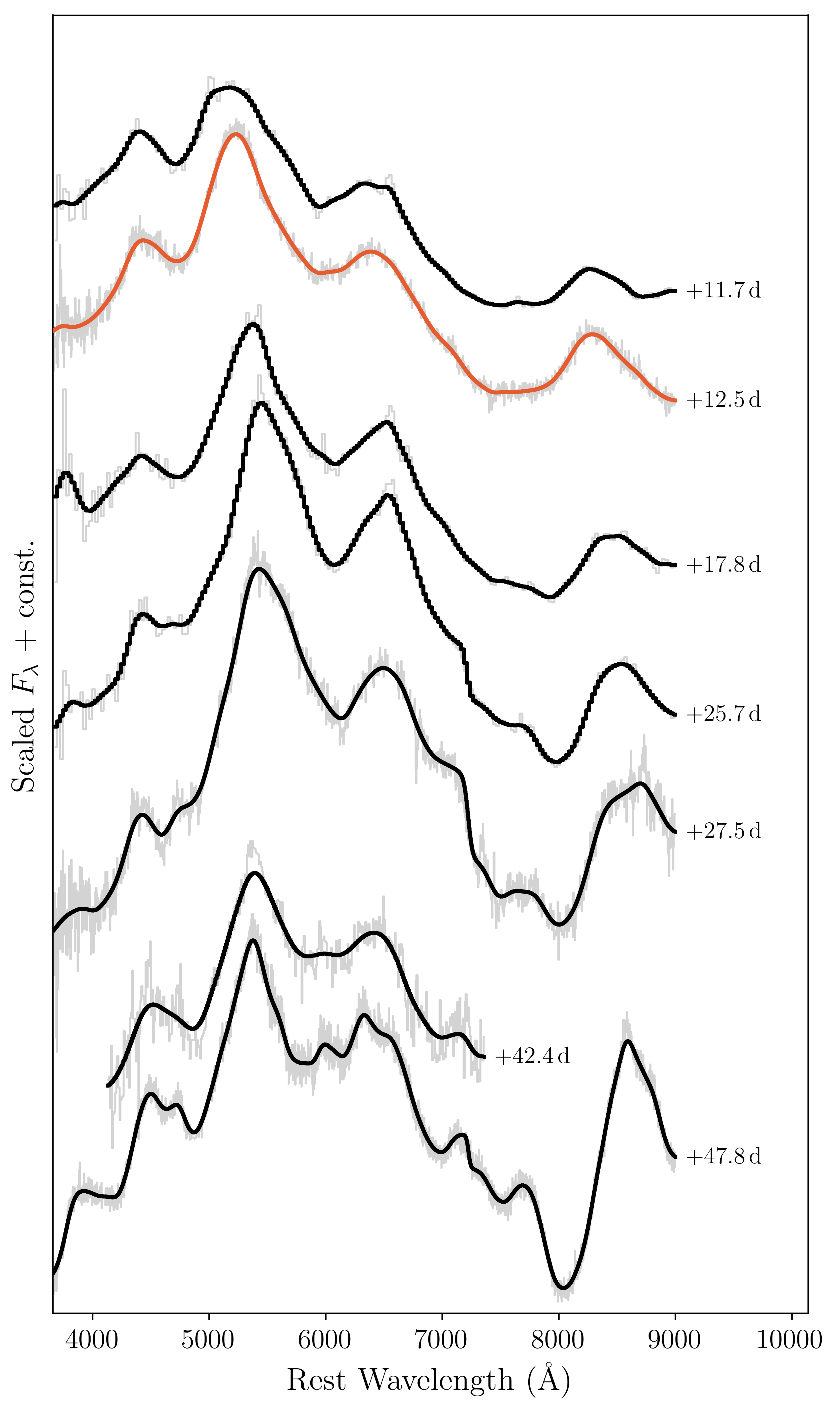}{0.47\textwidth}{}}
\caption{Optical spectra of SN\,2020bvc. Phase is relative to $t_0$, defined in \S\ref{sec:ztf-detection} as the time of last non-detection by ATLAS. The first spectrum is dominated by a blue continuum. By $\Delta t=5.7\,\days$ the spectrum strongly resembled a Ic-BL SN.
The raw spectrum is shown in light grey, and a smoothed spectrum (with host emission lines removed) is overlaid in black.
Spectra highlighted in orange are plotted compared to LLGRB-SNe at similar phases in Figure~\ref{fig:spec-comparison}.
\label{fig:spec}}
\end{figure*}

Epochs of spectroscopic observations are marked with
`S' in Figure~\ref{fig:full-lc},
and observation details are provided in Table~\ref{tab:SN2020bvc-spectra}.
The spectral sequence is shown in Figure~\ref{fig:spec},
and discussed in more detail in \S\ref{sec:analysis-spec}.
Both raw and smoothed versions of the spectra will be made available on WISeREP.

\begin{deluxetable}{lrrrr}[htb!]
\tablecaption{Spectroscopic observations of SN\,2020bvc. Epochs given since $t_0$ as defined in \S\ref{sec:ztf-detection}.
Velocities are derived from \ion{Fe}{2} absorption features as described in \S\ref{sec:vel}. \label{tab:SN2020bvc-spectra}} 
\tablewidth{0pt} 
\tablehead{\colhead{Date} & \colhead{$\Delta t$} & \colhead{Tel.+Instr.} & \colhead{Exp. Time} & \colhead{$v_\mathrm{ph}$} \\ 
\colhead{(UTC)} & \colhead{(d)} & & \colhead{(s)} & ($10^{4}\,\km\,\psec$)} 
\tabletypesize{\scriptsize} 
\startdata 
Feb 04.43 & 0.76 & P60+SEDM & 1800 & -- \\
Feb 07.36 & 3.7 & P60+SEDM & 1800 & $5.1\pm0.1$ \\
Feb 08.24 & 4.6 & LT+SPRAT & 600 & $2.58\pm0.51$ \\
Feb 09.36 & 5.7 & P60+SEDM & 1800 & -- \\
Feb 12.51 & 8.8 & P200+DBSP & 600 & $1.83\pm0.32$ \\
Feb 13.33 & 9.7 & P60+SEDM & 1800 & -- \\
Feb 15.33 & 11.7 & P60+SEDM & 1800 & -- \\
Feb 16.14 & 12.5 & NOT+ALFOSC & 1200 & $1.90\pm0.25$ \\
Feb 21.43 & 17.7 & P60+SEDM & 1800 & -- \\
Feb 29.42 & 25.8 & P60+SEDM & 1800 & -- \\
Mar 02.14 & 27.5 & NOT+ALFOSC & 1200 & -- \\
Mar 17.19 & 42.6 & LT+SPRAT & 900 & $1.72\pm0.32$ \\
Mar 22.50 & 47.9 & Keck1+LRIS & 300 & $1.79\pm0.39$ \\
\enddata 
\end{deluxetable}

\subsection{UV and X-ray Observations}
\label{sec:xray}

We obtained ten observations of SN\,2020bvc\footnote{The target name was PTF13ast, a previous SN hosted by UGC\,09379.} with the UV/optical (UVOT; \citealt{Roming2005}) and X-Ray Telescope (XRT; \citealt{Burrows2005}) on board
the Neil Gehrels \emph{Swift} observatory \citep{Gehrels2004}
under a target-of-opportunity program (PI: Schulze).
The first observation was on Feb 05.02 ($\Delta t=1.35$).
We also obtained two 10\,ks observations with the \chandra\ X-ray Observatory under Director's Discretionary Time (PI A. Ho), one epoch on Feb 16\footnote{ObsId 23171, \dataset [ADS/Sa.CXO\#obs/23171]} ($\Delta t=13.2$) and one epoch on Feb 29\footnote{ObsId 23172, \dataset [ADS/Sa.CXO\#obs/23172]} ($\Delta t=25.4$).

UVOT photometry was performed using the task \texttt{UVOTsource} in HEASoft\footnote{\url{http://heasarc.gsfc.nasa.gov/ftools}} version 6.25 \citep{Blackburn1999},
with a 3\arcsec-radius aperture.
For host subtraction, a template was constructed
from data prior to 2014 Dec 09.
Host-subtracted, Milky Way extinction-corrected values are provided in Table~\ref{tab:uvot-phot} in Appendix~\ref{sec:appendix-phot-table}.
XRT data were reduced using the online tool\footnote{\url{https://www.swift.ac.uk/user_objects/}} from the \emph{Swift} team \citep{Evans2007,Evans2009},
with the data binned by observation and centroiding turned off.
\chandra\ data were reduced using
the Chandra Interactive Analysis of Observations (CIAO) software package (v4.12; \citealt{Fruscione2006}).

Stacking the first 2.2\,ks of XRT observations (four nightly 0.6\,ks exposures) we detected 4 counts with an expected background of 0.16 counts.
The resulting count rate is $(2.9_{-1.9}^{+3.3}) \times 10^{-3}\,\psec$ (90\% confidence interval).
To convert count rate to flux,
we used a hydrogen column density $n_H = 9.90 \times 10^{19}\pcmsq$ \citep{HI4PI2016} and
a photon power-law index of $\Gamma=2$.
The resulting unabsorbed 0.3--10\,\kev\ flux is
$(9.3^{+10.6}_{-6.1}) \times 10^{-14}\,\erg\,\psec\,\pcmsq$,
and the luminosity is
$(1.4^{+1.7}_{-0.9}) \times 10^{41}\,\erg\,\psec$.
From prior \swift\ observations of the position of SN\,2020bvc,
we measured a 90\% upper limit of $<7.8 \times 10^{-4}\,$\,\psec, suggesting that the emission is not from the host.
We note that there is a discrepancy between our \swift\ measurements and those in \citet{Izzo2020},
who find a significantly higher XRT flux value.

In the first epoch of our \chandra\ observations,
a total of eight counts were detected in a 1\arcsec-radius region centered on the source.
To measure the background, we set an annulus around the source with an inner radius of 3\arcsec\ and an outer radius of 10\arcsec.
The average background was 0.21\,$\mathrm{arcsec}^{-2}$,
so the expected number of background counts within the source region is 0.65.
The 90\% confidence interval for the number of detected counts from the source is 3.67--13.16 \citep{Kraft1991}, so we conclude that the detection is significant.

We used CIAO to convert the count rate from the first observation ($(5.9^{+5.1}_{-3.3})\times10^{-4}$\,\psec) to flux,
assuming the same photon index and $n_H$ value as for the \swift\ observations,
finding an unabsorbed 0.5--7\,\kev\ flux of
$(7.2^{+6.3}_{-3.9}) \times 10^{-15}\,\erg\,\psec\,\pcmsq$.
In the second epoch,
seven counts were detected,
with a count rate of
$(5.9^{+5.1}_{-3.2})\times10^{-4}$\,\psec\ and an
unabsorbed 0.5--7\,\kev\ flux of
$(7.2^{+6.2}_{-4.0}) \times 10^{-15}\,\erg\,\psec\,\pcmsq$.
The corresponding luminosity is
$(1.1^{+1.0}_{-0.6}) \times 10^{40}\,\erg\,\psec$
in each epoch.
In \S\ref{sec:modeling-xray} we compare the X-ray light curve to that of other Ic-BL SNe.
Again, we note a discrepancy with the measurements of \citet{Izzo2020}, who find a significantly higher flux value than we do (as shown in their Fig.~2).

\subsection{Submillimeter and Radio Observations}

As listed in Table~\ref{tab:radio-flux},
we obtained eight observations of SN\,2020bvc\footnote{Program VLA/20A-374; PI A. Ho} with the Karl G. Jansky Very Large Array (VLA; \citealt{Perley2011}),
while the array was in C configuration.
3C286 was used as the flux density and bandpass calibrator and J1417+4607 as the complex gain calibrator.
Data were calibrated using the automated pipeline available in the Common Astronomy Software Applications (CASA; \citealt{McMullin2007}),
with additional flagging performed manually,
and imaged\footnote{Cell size was 1/10 of the synthesized beamwidth, field size was the smallest magic number
($10 \times 2^n$) larger than the number of cells needed to cover the primary beam.} using the CLEAN algorithm \citep{Hogbom1974}.

\begin{deluxetable}{lcccc}[htb!]
\tablecaption{Submillimeter- and centimeter-wavelength radio observations of SN\,2020bvc.
The 230\,\ghz\ measurement was obtained using the Submillimeter Array (upper limit given as 1$\sigma$ RMS) and the lower-frequency measurements were obtained using the Very Large Array.
\label{tab:radio-flux}} 
\tablewidth{0pt} 
\tablehead{ 
\colhead{Start Date} & \colhead{Time on-source} & \colhead{$\Delta t$} & \colhead{$\nu$} & \colhead{Flux Density} \\ 
\colhead{(UTC)} & \colhead{(hr)} & \colhead{(days)} & \colhead{(GHz)} & \colhead{$(\mu$Jy) }
} 
\tabletypesize{\scriptsize} 
\startdata 
Feb 09.42 & 4.7 & 5.8 & 230 & $<250$ \\
Feb 16.67 & 0.4 & 13.0 & 10 & $63\pm6$ \\
Feb 20.64 & 0.4 & 17.0 & 6 & $83\pm6$ \\
Feb 27.64 & 0.4 & 24.0 & 3 & $111\pm10$ \\
Mar 02.63 & 0.4 & 28.0 & 15 & $33\pm4$ \\
Mar 09.60 & 0.4 & 35.0 & 10 & $50\pm5$ \\
Mar 13.59 & 0.4 & 39.0 & 3 & $106\pm10$ \\
Mar 17.33 & 0.4 & 42.7 & 6 & $63\pm6$ \\
\enddata 
\end{deluxetable}

Motivated by the detection of LLGRB\,980425/SN\,1998bw \citep{Galama1998} at 2\,mm \citep{Kulkarni1998}
and of LLGRB\,171205A/SN\,2017iuk at 3\,mm and 1\,mm \citep{Perley2017-alma},
we also observed\footnote{Program 2019B-S026; PI A. Ho} SN\,2020bvc with the Submillimeter Array \citep{Ho2004}, which was in its compact configuration.\footnote{RxA and RxB receivers were tuned to a local-oscillator frequency of 223.556\,\ghz, providing 16\,\ghz\ of overlapping bandwidth: 211.56\,\ghz--219.56\,\ghz\ in the lower side-band and 227.56--235.56\,\ghz\ in the upper side-band with a spectral resolution of 140.0\,kHz per channel.}
The phase and amplitude gain calibrators were J1419+383 and J1310+323, the passband calibrator was 3C84, and the flux calibrator was Uranus.
Data were calibrated using the SMA MIR IDL package and imaged using MIRIAD \citep{Sault1995}.

\begin{figure}[htb!]
    \centering
    \includegraphics[width=\linewidth]{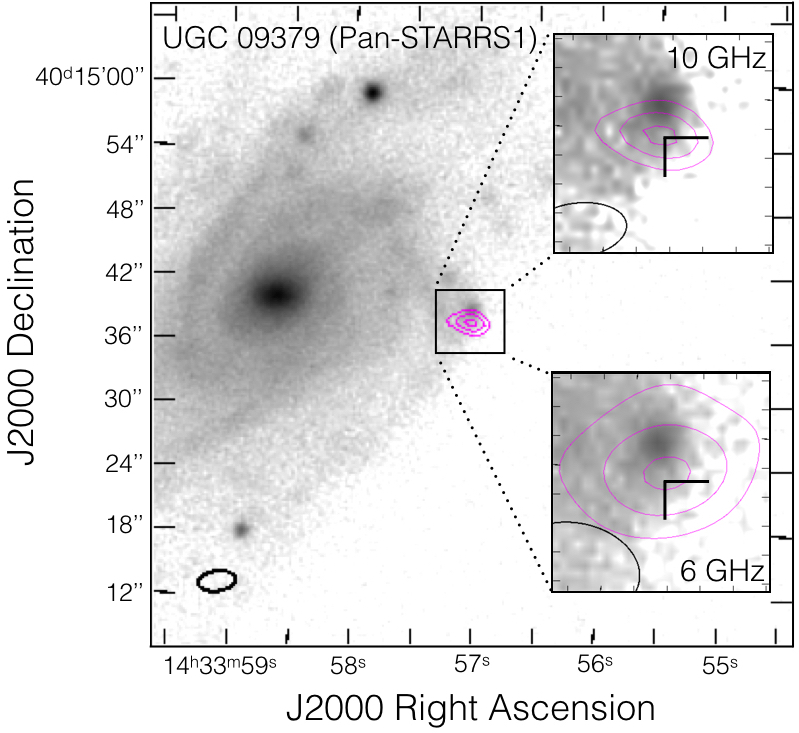}
    \caption{Image of the 10\ghz\ and 6\,\ghz\ VLA observations of SN2020bvc.
    The background image of UGC\,09379 is from Pan-STARRS1 \citep{Chambers2016,Flewelling2016}. The radio data is overlaid as contours and the size of the synthesized beam is shown as an ellipse on the bottom left. The position of the optical transient is shown as cross-hairs in the zoom-in panels.}
    \label{fig:radio-image}
\end{figure}

No source was detected by the SMA,
with a spectral channel-averaged 1$\sigma$ RMS of 0.25\,mJy.
A source was detected at the position of SN\,2020bvc in all epochs of VLA observations,
and no sources were detected elsewhere in the host galaxy.
Using the task \texttt{imfit},
we confirm that the radio source is a point source at all frequencies, and that the centroid is at the position of the optical transient.
In Figure~\ref{fig:radio-image} we show the centroid of the radio emission and the position of the optical transient,
and that both are offset from the nearby \ion{H}{2} region.

\begin{figure}[!htb]
    \centering
    \includegraphics[width=\linewidth]{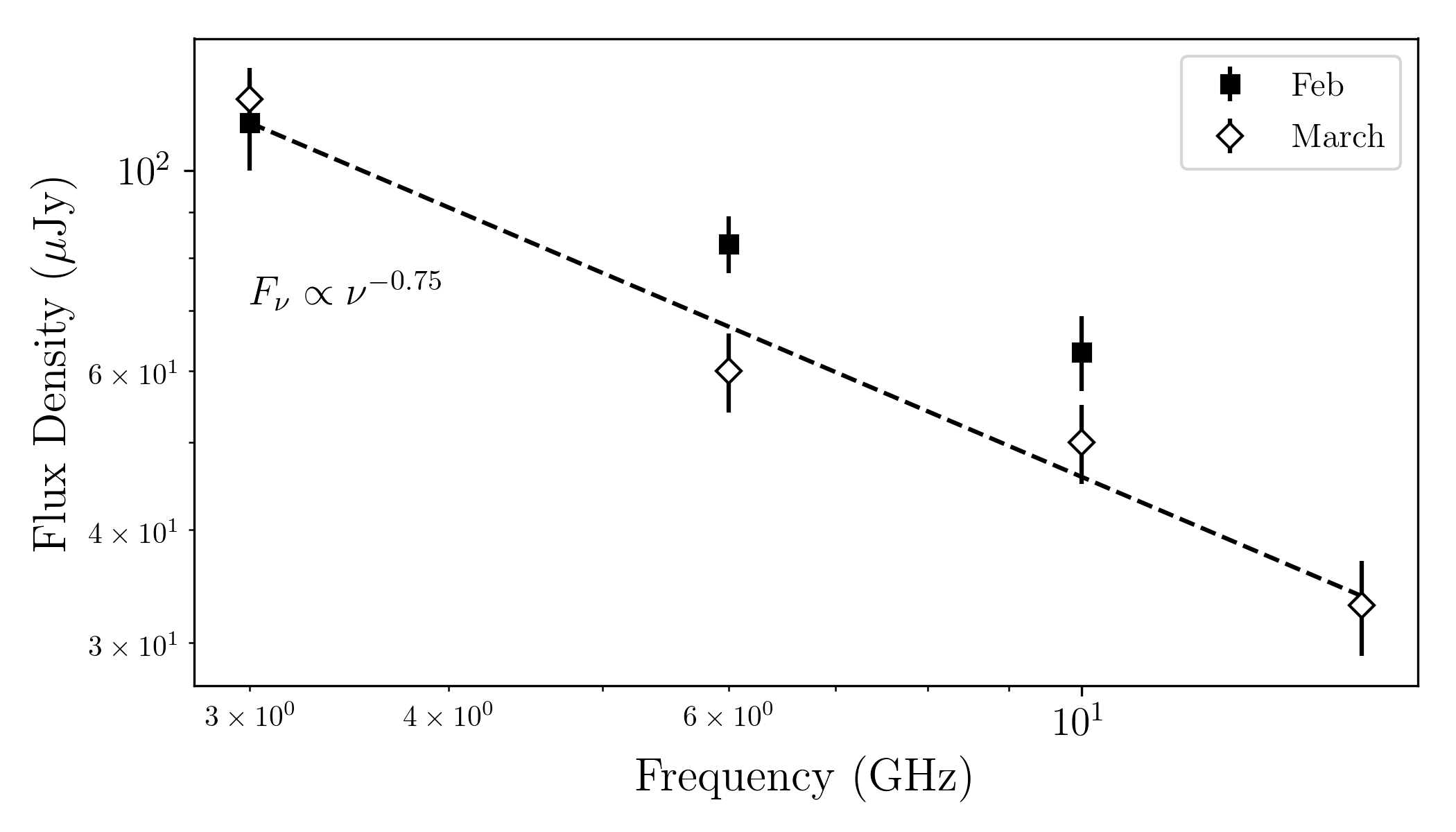}
    \caption{Radio SED of SN\,2020bvc from VLA observations spanning two months post-explosion.}
    \label{fig:radio-sed}
\end{figure}

In the first observation ($\Delta t=13\,\days$)
the 10\,\ghz\ peak flux density was $63\pm6\,\mu$Jy,
corresponding to a luminosity of
$1.0 \times 10^{27}\,\erg\,\psec\,\phz$ \citep{Ho2020-vla}.
The source was brighter at lower frequencies,
and there is marginal (2$\sigma$) evidence for fading
at 6\,\ghz\ ($F_\nu \propto t^{-0.23\pm0.15}$) and 10\,\ghz\ ($F_\nu \propto t^{-0.25\pm0.16}$),
but no evidence for fading at 3\,\ghz.
The radio SED is shown in Figure~\ref{fig:radio-sed}.
In \S\ref{sec:modeling-radio}
we compare the 10\,\ghz\ light curve to that of other Ic-BL SNe
and use the SED to model the forward shock.

\subsection{Search for a Gamma-ray Burst}
\label{sec:grb-search}

The third Interplanetary Network (IPN\footnote{\url{http://ssl.berkeley.edu/ipn3/index.html}}) consists of six spacecraft that provide all-sky full-time monitoring for high-energy bursts.
The most sensitive detectors in the IPN are the \emph{Swift} Burst Alert Telescope (BAT; \citealt{Barthelmy2005}) the \emph{Fermi} 
Gamma-ray Burst Monitor (GBM; \citealt{Meegan2009}),
and the Konus instrument on the WIND spacecraft \citep{Aptekar1995}.

We searched the \emph{Fermi} GBM Burst Catalog\footnote{\url{https://heasarc.gsfc.nasa.gov/W3Browse/fermi/fermigbrst.html}} \citep{Gruber2014,vonKienlin2014,Bhat2016},
the \fermi-GBM Subthreshold Trigger list\footnote{\url{https://gcn.gsfc.nasa.gov/fermi\_gbm\_subthresh\_archive.html}} (with reliability flag \texttt{!=2}), the
\swift\ GRB Archive\footnote{\url{https://swift.gsfc.nasa.gov/archive/grb\_table/}}, the IPN master list\footnote{\url{http://ipn3.ssl.berkeley.edu/masterli.txt}},
and the Gamma-Ray Coordinates Network archives\footnote{\url{https://gcn.gsfc.nasa.gov/gcn3\_archive.html}} for a GRB between the last ZTF non-detection (Feb 02.56) and the first ZTF detection (Feb 04.34).
The only bursts consistent with the position of SN\,2020bvc were detected by Konus-WIND, but likely arose from an X-ray binary system that was active at the time.
We conclude that SN\,2020bvc had no detected GRB counterpart.

Given the lack of a detected GRB,
we can use the sensitivity of the IPN spacecraft to set a limit on the isotropic-equivalent gamma-ray luminosity $L_{\gamma,\mathrm{iso}}$ of any counterpart.
During the time interval of interest (16\,h)
the position of SN\,2020bvc was within the coded field-of-view of the BAT for only 5.25 hours\footnote{Search conducted using \url{https://github.com/lanl/swiftbat\_python}}. So, we cannot set a useful limit using BAT.

\fermi/GBM had much better coverage\footnote{Search conducted using \url{https://github.com/annayqho/HE\_Burst\_Search}},
with the position of SN\,2020bvc visible most of the time (12.7\,h).
\fermi/GBM is in a low-Earth ($\sim1.5\,$hr) orbit,
and the position was occulted by the Earth for ten minutes per orbit, although
in six out of ten of these occultations the position was visible to \swift/BAT.
During the interval of interest \fermi\ went through
five South Atlantic Anomaly passages ranging from 10--30\,min in duration.
Since SN\,2020bvc was visible to GBM most of the time,
it is worthwhile to use the GBM sensitivity to place a limit on an accompanying GRB.
For a long and soft template\footnote{a smoothly broken power law with low-energy index $-1.9$ and high-energy index $-2.7$, and $E_\mathrm{pk}=70\,\kev$}
the GBM sensitivity is a few $\times 10^{-8}\,\erg\,\psec\,\pcmsq$ (see the discussion in \S 2.7 of \citealt{Ho2019gep}),
so the isotropic equivalent luminosity $L_\mathrm{\gamma,\mathrm{iso}} \lesssim \mathrm{few} \times 10^{46}\,\erg\,\psec$.

We obtain our most conservative lower limit on accompanying GRB emission using Konus-WIND,
which had continuous visibility of the SN\,2020bvc position due to its position beyond low-Earth orbit.
Assuming a Band spectral model with $\alpha=-1$, $\beta=-2.5$, and $E_\mathrm{pk}=50$--500\,keV,
the limiting 20--1500\,keV peak energy flux for a 2.944\,s timescale is $1$--2$\times 10^{47}\,$\erg\,\psec,
corresponding to an upper limit on the peak isotropic-equivalent gamma-ray luminosity of 1.7--3.4$\times 10^{47}\,\erg\,\psec$.

For reference, classical GRBs have $L_{\gamma,\mathrm{iso}}>10^{49.5}\,\erg\,\psec$ \citep{Cano2017}.
LLGRBs have $L_{\gamma,\mathrm{iso}}<10^{48.5}\,\erg\,\psec$: LLGRB\,060218 had $L_{\gamma,\mathrm{iso}}=2.6 \times 10^{46}\,\erg\,\psec$ \citep{Cano2017}.
However, GBM would be unlikely to detect a GRB like LLGRB\,060218 accompanying SN\,2006aj because of the low peak energy $E_\mathrm{pk}\sim5\,\kev$ and long duration $T_{90}\sim2100\,$s \citep{Cano2017}.
Weak signals longer than 100\,s look like background evolution to GBM because the detector background can change significantly over 100--200\,s.
Therefore, although a classical GRB is clearly ruled out (both by the lack of GRB and the lack of strong afterglow emission) we cannot rule out the possibility that SN\,2020bvc had prompt emission identical to an LLGRB like 060218. We revisit the difficulty of ruling out an LLGRB in \S\ref{sec:summary}.

\section{Light Curve Analysis}
\label{sec:lc-analysis}

\subsection{Comparisons to Other Ic-BL SNe}
\label{sec:lc-comparisons}

The P48 light curve of SN\,2020bvc is shown in Figure~\ref{fig:full-lc}, aligned with the light curve of SN\,2006aj.
The relative time of LLGRB\,060218 is close to the time of the ATLAS non-detection, supporting our choice of the ATLAS non-detection as our estimated epoch of first light $t_0$.
In both SN\,2006aj and SN\,2020bvc, the first peak fades on the timescale of one day, followed by the rise of the main peak of the light curve.
In \S\ref{sec:shock-cooling} we model the first peak as arising from shock-cooling of extended low-mass material
and discuss the implication of the fact that SN\,2006aj and SN\,2020bvc have similar first peaks.

\begin{figure*}[htb!]
    \centering
    \includegraphics[width=\textwidth]{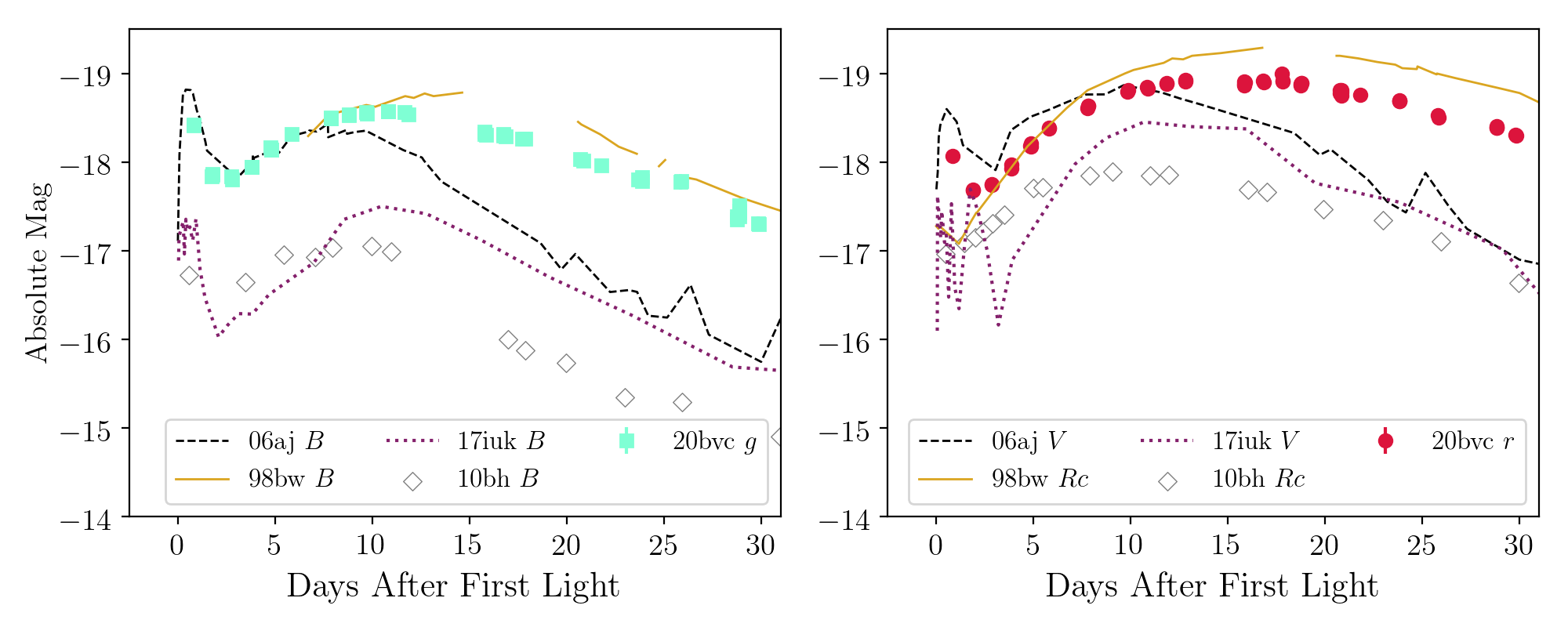}
    \caption{Comparison of the light curve of SN\,2020bvc to nearby LLGRB-SNe, shifted to a common redshift. The SN\,1998bw light curve was taken from Table~2 of \citet{Clocchiatti2011}, which uses data from \citet{Galama1998} and \citet{Sollerman2002}, and corrected for Milky Way extinction. The SN\,2006aj light curve was taken from the Open SN catalog and corrected for MW extinction, with original data from \citet{Modjaz2006}, \citet{Bianco2014}, and \citet{Brown2014}. The SN\,2010bh data were taken as-is from \citet{Cano2011}. The SN\,2017iuk data were taken from \citet{Delia2018} and corrected for MW extinction.}
    \label{fig:lc-comparison}
\end{figure*}

The second peak has a rise time from first light of
13--15\,\days\ in $r$-band ($M_{r,\mathrm{pk}}=-18.7\,$mag) and 10--12\,\days\ in $g$-band ($M_{g,\mathrm{pk}}=-18.3\,$mag).
In Figure~\ref{fig:lc-comparison} we compare the light curve to several LLGRB-SNe.
The timescale of the second peak is most similar to that of SN\,1998bw in $r$-band and most similar to SN\,2006aj and SN\,2017iuk in $g$-band.
The peak luminosity is intermediate to SN\,2006aj and SN\,1998bw.
We discuss the implications in \S\ref{sec:nickel},
when we use the light curve of the main peak to measure properties of the explosion such as the nickel mass, ejecta mass, and kinetic energy.

\subsection{Blackbody Fits}
\label{sec:bbfits}

We fit blackbody functions to the photometry of SN\,2020bvc in order to measure the evolution of the bolometric luminosity $L_\mathrm{bol}$, photospheric radius $R_\mathrm{ph}$, and effective temperature $T_\mathrm{eff}$.
First we manually selected 23 time bins as close as possible to epochs with observations in multiple filters.
We binned the P48 light curve such that observations in a single band clustered within a few hours were averaged together.
For each time bin, we constructed an SED by linearly interpolating the UV and optical light curves shown in Figure~\ref{fig:lc-grid}.
After $\Delta t=2\,$d we exclude the UVW2 point in the fits, because it shows an excess compared to the blackbody function.
For each SED, we used the nonlinear least squares routine of \texttt{curve\_fit} in \texttt{scipy} \citep{Virtanen2020} to fit a blackbody.
To estimate uncertainties,
we performed a Monte Carlo simulation with 600 realizations of the data.
The size of the error bar on each point
is a 30\% fractional systematic uncertainty,
chosen to obtain a combined $\chi^2/\mathrm{dof} \approx 1$ across all epochs.

\begin{figure*}[htb!]
    \centering
    \includegraphics[width=\textwidth]{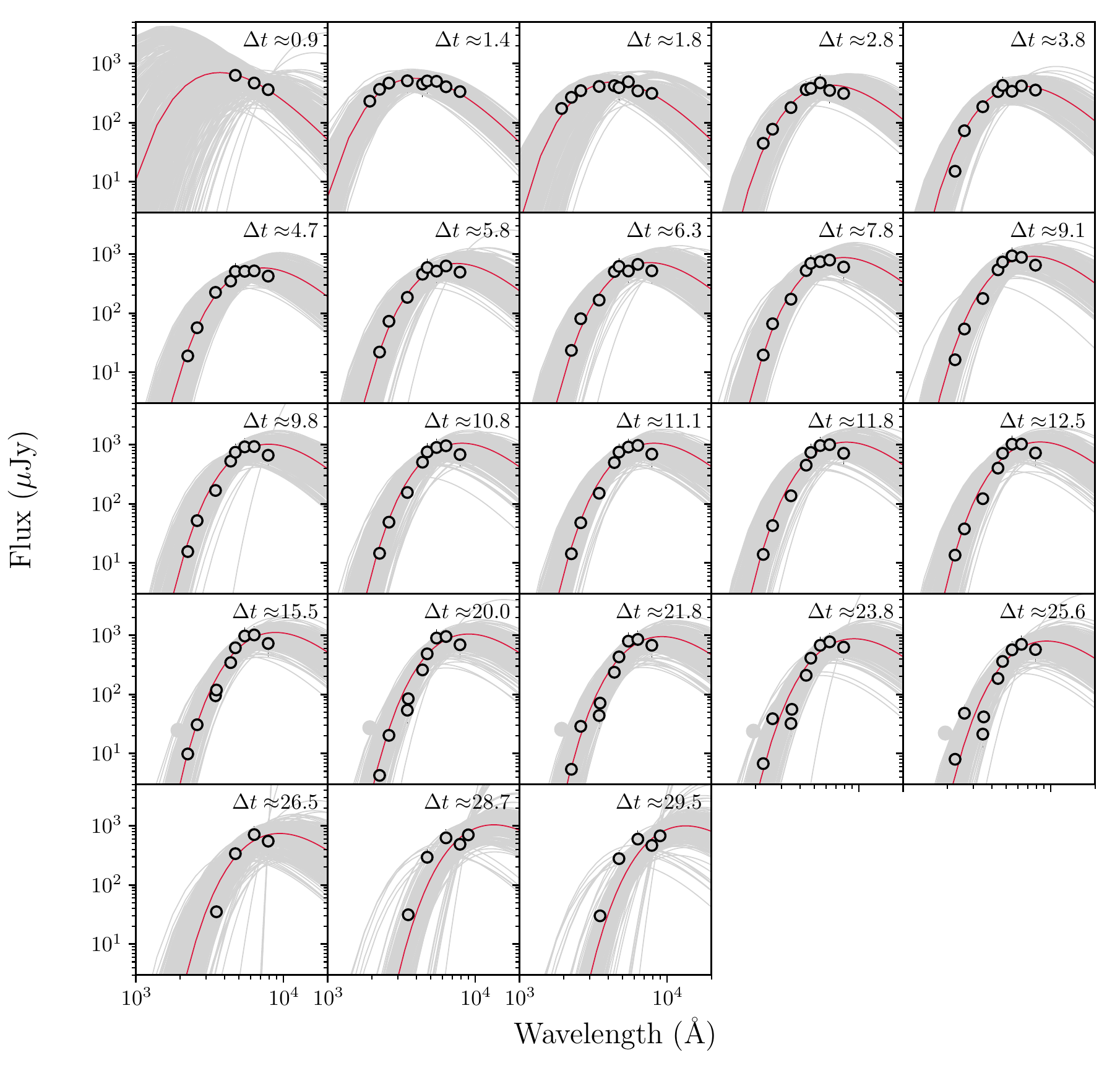}
    \caption{Blackbody fits to optical and \emph{Swift}/UVOT photometry of SN\,2020bvc.
    Photometry has been interpolated onto common epochs as described in \S\ref{sec:bbfits}.
    Fit was run through a Monte Carlo with 600 realizations of the data. Individual fits are shown as thin grey lines; dispersion corresponds to overall uncertainties in the fits. Only outlined points were included in the fits.}
    \label{fig:bbfits}
\end{figure*}

\begin{figure}[htb!]
    \centering
    \includegraphics[width=0.86\linewidth]{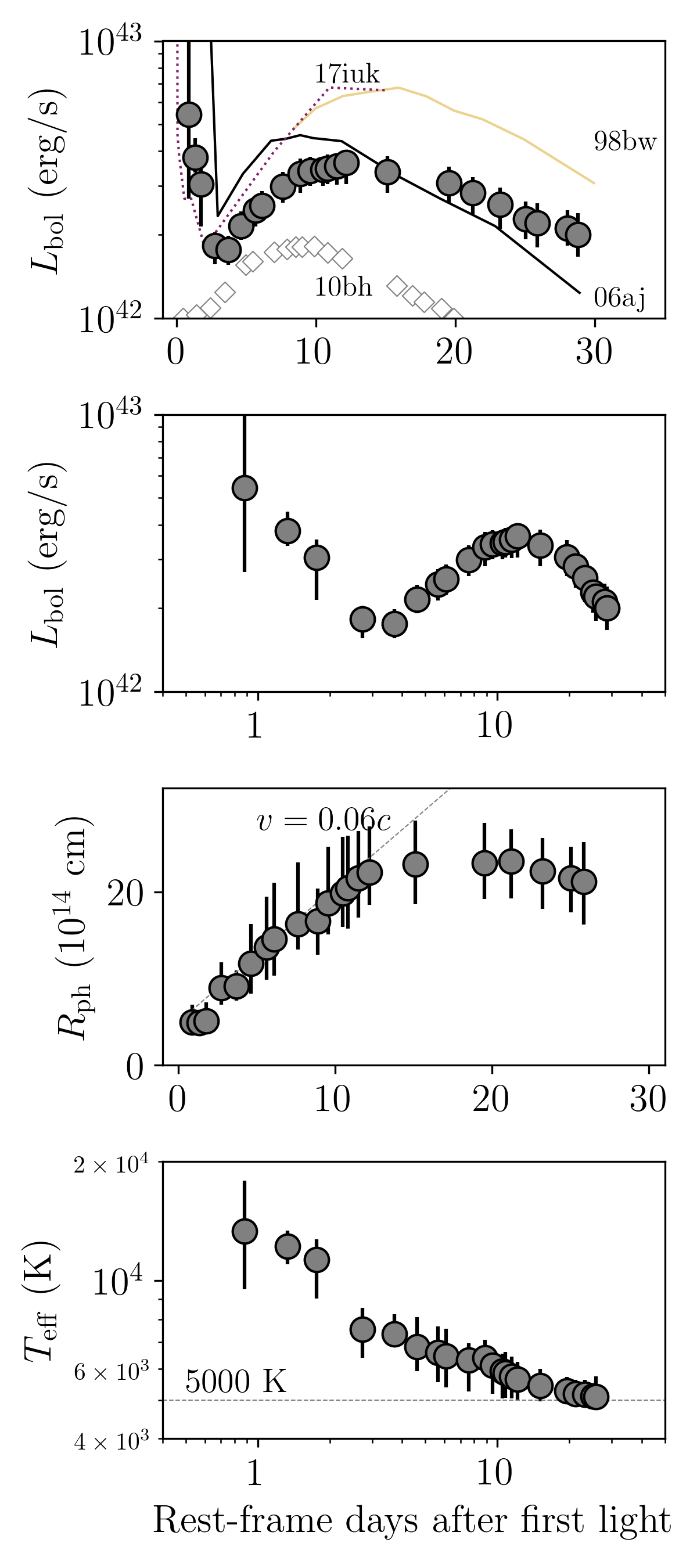}
    \caption{Blackbody evolution of SN\,2020bvc.
    Top panel: bolometric light curve compared to LLGRB-SNe: SN\,2006aj and SN\,1998bw
    (UBVRI; \citealt{Cano2013}), SN\,2010bh (BVRI; \citealt{Cano2013}), SN\,2017iuk (spectral modeling; \citealt{Izzo2019}).
    We add early $L_\mathrm{bol}$ measurements of SN\,2006aj from \citet{Campana2006}.
    Second panel: bolometric light curve in log-log space.
    Third panel: photospheric radius, with a dotted line indicating $v=18,000\km\,\psec$.
    Bottom panel: effective temperature, with a horizontal line marking 5000\,K, the recombination temperature of carbon and oxygen.}
    \label{fig:bb-evolution}
\end{figure}

The fits are shown in Figure~\ref{fig:bbfits}.
The best-fit parameters are listed in Table~\ref{tab:bb-evolution}
and plotted in Figure~\ref{fig:bb-evolution}.
$L_\mathrm{bol}$ peaks after $\Delta t\approx12$--14\,\days\
at $L_\mathrm{bol,pk}=4 \times 10^{42}\,\erg\,\psec$,
and $R_\mathrm{ph}$ increases by $v_\mathrm{ph} \approx 0.06\,c$, which is consistent with the 18,000\,\km\,\psec\ that we measure from the spectra in \S\ref{sec:vel}.
Using trapezoidal integration we find a total radiated energy $E_\mathrm{rad}=7.1 \times 10^{48}\,\erg$.

In the top panel of Figure~\ref{fig:bb-evolution} we show the
evolution of $L_\mathrm{bol}$ compared to nearby LLGRB-SNe.
To make the comparison, we chose bolometric light curves constructed using similar filters: UBVRI for the second peak of SN\,2006aj and SN\,1998bw, and BVRI for SN\,2010bh \citep{Cano2013}.
We could not find a similar bolometric light curve for the second peak of SN\,2017iuk, so we used $L_\mathrm{bol}$ from the spectral modeling of \citet{Izzo2019} and caution that this is not a direct comparison.
For SN\,2006aj we used an early measurement of the bolometric luminosity from the UVOT data \citep{Campana2006}.
SN\,2020bvc and SN\,2017iuk have a similarly fast-declining first peak;
early detailed UV observations were not obtained for SN\,1998bw and SN\,2010bh.
Overall, SN\,1998bw is the most luminous LLGRB-SN,
followed by SN\,2006aj and SN\,2020bvc, which are similar to one another.
SN\,2010bh is significantly less luminous.
We revisit these comparisons when we calculate the explosion parameters of SN\,2020bvc in \S\ref{sec:modeling-opt-lc}.

\begin{deluxetable}{lrrrr}[htb!]
\tablecaption{Blackbody evolution of SN\,2020bvc. The first epoch is from fitting the optical spectrum (\S\ref{sec:analysis-spec}). The remaining epochs are from fitting multi-band photometry (\S\ref{sec:bbfits}). \label{tab:bb-evolution}} 
\tablewidth{0pt} 
\tablehead{\colhead{$\Delta t$} &
\colhead{$L_\mathrm{bol}$} & \colhead{$T_\mathrm{eff}$} & \colhead{$R_\mathrm{ph}$}
\\ 
\colhead{(d)} & \colhead{($10^{42}\,\erg\,\psec$)} & \colhead{($10^3$\,K)} & ($10^{14}\,\cm$)} \tabletypesize{\scriptsize} 
\startdata 
0.67 & $5.6\pm0.3$ & $13.2\pm0.3$ & $5.1\pm0.1$ \\
0.9 & $5.4^{+6.2}_{-2.7}$ & $13.3^{+4.6}_{-3.8}$ & $5.0^{+2.0}_{-1.2}$ \\
1.4 & $3.8^{+0.7}_{-0.4}$ & $12.2^{+1.2}_{-1.2}$ & $4.9^{+0.9}_{-0.7}$ \\
1.8 & $3.1^{+0.5}_{-0.9}$ & $11.3^{+1.4}_{-2.3}$ & $5.1^{+2.1}_{-0.9}$ \\
2.8 & $1.8^{+0.2}_{-0.3}$ & $7.6^{+1.0}_{-1.2}$ & $8.9^{+2.9}_{-1.9}$ \\
3.8 & $1.8^{+0.2}_{-0.2}$ & $7.4^{+0.9}_{-0.6}$ & $9.1^{+1.8}_{-1.7}$ \\
4.7 & $2.1^{+0.3}_{-0.2}$ & $6.8^{+1.3}_{-0.9}$ & $11.7^{+4.6}_{-3.5}$ \\
5.8 & $2.4^{+0.3}_{-0.3}$ & $6.6^{+1.1}_{-1.0}$ & $13.6^{+5.8}_{-3.8}$ \\
6.3 & $2.5^{+0.3}_{-0.3}$ & $6.5^{+1.1}_{-1.1}$ & $14.5^{+6.6}_{-4.2}$ \\
7.8 & $3.0^{+0.4}_{-0.4}$ & $6.3^{+0.7}_{-1.0}$ & $16.3^{+7.1}_{-3.0}$ \\
9.1 & $3.3^{+0.4}_{-0.5}$ & $6.4^{+0.7}_{-0.5}$ & $16.6^{+3.8}_{-3.9}$ \\
9.8 & $3.4^{+0.4}_{-0.4}$ & $6.1^{+0.6}_{-0.9}$ & $18.7^{+6.5}_{-3.6}$ \\
10.8 & $3.4^{+0.4}_{-0.4}$ & $5.9^{+0.6}_{-0.9}$ & $19.9^{+6.5}_{-3.9}$ \\
11.1 & $3.5^{+0.4}_{-0.4}$ & $5.9^{+0.8}_{-0.8}$ & $20.5^{+6.1}_{-4.7}$ \\
11.8 & $3.5^{+0.4}_{-0.5}$ & $5.7^{+0.7}_{-0.7}$ & $21.6^{+5.4}_{-4.5}$ \\
12.5 & $3.6^{+0.4}_{-0.6}$ & $5.6^{+0.6}_{-0.6}$ & $22.2^{+5.3}_{-3.7}$ \\
15.5 & $3.4^{+0.5}_{-0.5}$ & $5.4^{+0.6}_{-0.5}$ & $23.2^{+5.0}_{-4.6}$ \\
20.0 & $3.1^{+0.4}_{-0.5}$ & $5.3^{+0.4}_{-0.4}$ & $23.3^{+4.6}_{-4.1}$ \\
21.8 & $2.8^{+0.4}_{-0.5}$ & $5.2^{+0.4}_{-0.3}$ & $23.5^{+3.7}_{-4.3}$ \\
23.8 & $2.6^{+0.4}_{-0.5}$ & $5.2^{+0.5}_{-0.3}$ & $22.4^{+3.8}_{-4.4}$ \\
25.6 & $2.3^{+0.3}_{-0.4}$ & $5.1^{+0.4}_{-0.3}$ & $21.6^{+3.6}_{-3.9}$ \\
26.5 & $2.2^{+0.4}_{-0.4}$ & $5.1^{+0.6}_{-0.3}$ & $21.2^{+4.6}_{-5.0}$ \\
28.7 & $2.1^{+0.3}_{-0.3}$ & $3.6^{+0.2}_{-0.2}$ & $40.7^{+7.3}_{-4.1}$ \\
29.5 & $2.0^{+0.4}_{-0.3}$ & $3.6^{+0.2}_{-0.2}$ & $40.6^{+6.9}_{-4.9}$ \\
\enddata 
\end{deluxetable}

\section{Spectroscopic Properties}
\label{sec:analysis-spec}

\subsection{Spectroscopic Evolution and Comparisons}
\label{sec:spec-evol-comp}

As outlined in \S\ref{sec:spec}, we obtained 13 spectra of SN\,2020bvc in the 50 days following discovery, shown in Figure~\ref{fig:spec}.
Here we discuss the spectroscopic evolution in more detail and compare it to LLGRB-SNe.

\begin{figure*}[hbt!]
    \centering
    \includegraphics[width=0.85\textwidth]{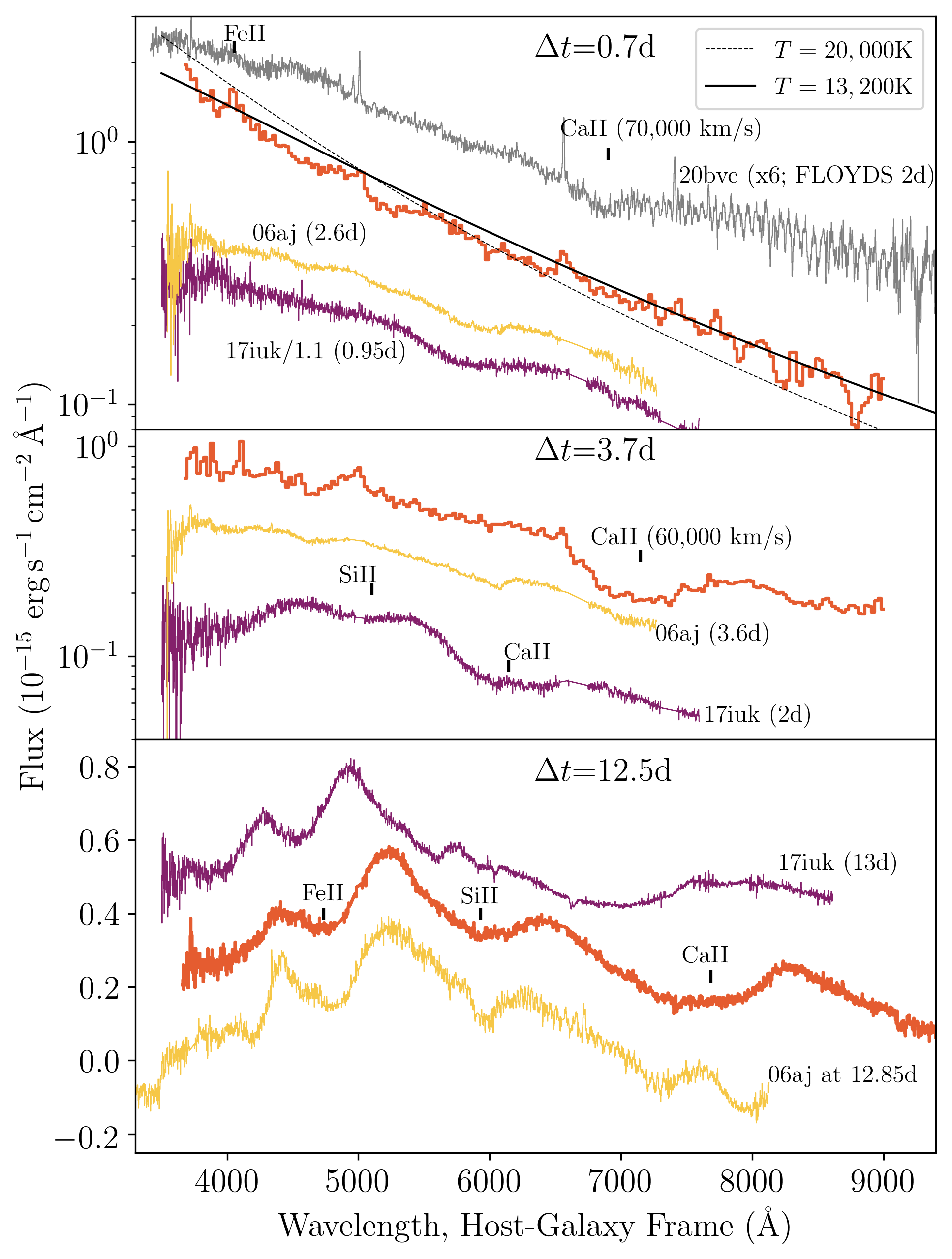}
    \caption{
    Spectra of SN\,2020bvc compared to spectra of LLGRB\,171205A/SN\,2017iuk (from \citealt{Izzo2019} and LLGRB\,060218/SN\,2006aj (from \citealt{Fatkhullin2006,Modjaz2006}/WISeREP) at similar epochs.
    In the top panel,
    we also show the blackbody fits described in
    \S\ref{sec:analysis-spec},
    and the spectrum of SN\,2020bvc at $\Delta t=1.9\,\days$ downloaded from WISeREP \citep{Hiramatsu2020} and obtained by the FLOYDS-N instrument on Faulkes Telescope North \citep{Brown2013}. The identification of \ion{Fe}{2} and \ion{Ca}{2} at 70,000\,\km\,\psec\ is from \citet{Izzo2020}.
    }
    \label{fig:spec-comparison}
\end{figure*}

The first spectrum ($\Delta t=0.7\,\days$) is shown in the top panel of Figure~\ref{fig:spec-comparison},
together with two blackbody fits.
The spectrum is best described
by a blackbody with
$L_\mathrm{bol}=(5.62\pm0.25) \times 10^{42}\,\erg\,\psec$,
$T_\mathrm{eff}=(13.21\pm0.27)\times10^{3}\,\kelvin$,
and
$R_\mathrm{ph}=(5.09\pm0.10) \times 10^{14}\,\cm$.
Here we are reporting statistical errors on the fit,
but there is also considerable systematic uncertainty due to being on the Rayleigh-Jeans tail.
We repeated the fit fixing $T_\mathrm{eff}=20,000\,\kelvin$,
and found $R=3.4 \times10^{14}\,\cm$ and $L=1.3\times10^{43}\,\erg\,\psec$.
Assuming that the value of
$R_\mathrm{ph}\approx 5 \times 10^{14}\,\cm$ at $\Delta t=0.7\,\days$ is much larger than the value of $R_\mathrm{ph}$ at $t_0$, we can estimate that the mean velocity until 0.7\,d is $5 \times 10^{14}\,\cm/0.7\,\days=0.3\,c$.
Taking the last ZTF non-detection as $t_0$, the mean velocity is reduced to $5 \times 10^{14}\,\cm/1.8\,\days=0.1\,c$.

For comparison, in the top panel of Figure~\ref{fig:spec-comparison} we show a higher-resolution spectrum obtained at $\Delta t=1.9$ and presented in \citet{Izzo2020}.
We mark the
\ion{Fe}{2} and \ion{Ca}{2} at $v_\mathrm{exp}=70,000\,\km\,\psec$ that 
\citet{Izzo2020} identified in their analysis,
which are not clearly distinguishable in the SEDM spectrum.
We also show early spectra of LLGRB-SNe:
a spectrum of LLGRB\,171205A/SN\,2017iuk at $\Delta t=1.5\,$hr from \citet{Izzo2019} and a spectrum of LLGRB\,060218/SN\,2006aj at $\Delta t=2.6\,\days$ from \citet{Fatkhullin2006}.
Both spectra are dominated by continuum, with a broad absorption feature near 5900\,\AA\ that is not clearly seen in the early spectrum of SN\,2020bvc.

The next spectrum of SN\,2020bvc was obtained at $\Delta t=3.7\,\days$,
which we show in the middle panel of Figure~\ref{fig:spec-comparison}.
A broad absorption feature is present at 7300\,\AA,
which in Figure~\ref{fig:spec} appears to shift redward with time.
For comparison, and to assist with identification of this feature, we compare the spectrum to two LLGRB-SN spectra obtained at a similar epoch,
LLGRB\,060218/SN\,2006aj \citep{Fatkhullin2006} and LLGRB\,171205A/SN\,2017iuk \citep{Izzo2019}.
The spectrum of SN\,2020bvc most closely resembles that of SN\,2017iuk.
We show two features in the SN\,2017iuk spectrum identified by \citet{Izzo2019},
\ion{Ca}{2} and \ion{Si}{2} at very high velocities
(105,000\,\km\,\psec\ for \ion{Ca}{2}).
Based on the similarity between the spectra, we also attribute the broad absorption feature to \ion{Ca}{2}.
To measure the expansion velocity we measure the minimum of the absorption trough,
finding $v_\mathrm{exp}=60,000\,\km\,\psec$ (based on the Gaussian center) and
a full-width at half-maximum (FWHM) of $0.16\,c$, or 48,000\,\km\,\psec.
The spectrum of SN\,2006aj shows hints of broad absorption features at similar wavelengths, but the lack of coverage on the red side makes it difficult to confirm the \ion{Ca}{2} absorption.

After $3.7\,\days$, the spectra of SN\,2020bvc can be readily classified as Type Ic-BL.
A spectrum of SN\,2020bvc near peak optical light ($\Delta t=13\,\days$) is shown in the bottom panel of Figure~\ref{fig:spec-comparison} compared to SN\,2006aj and SN\,2017iuk at a similar epoch.
The \ion{Si}{2} and \ion{Ca}{2} absorption lines are clearly broader in the spectrum of SN\,2020bvc than in the spectrum of SN\,2006aj,
although the centroids are at a similar wavelength,
suggesting that the expansion velocities are similar.
The absorption lines are at a higher expansion velocity in the spectrum of SN\,2017iuk than in the spectrum of SN\,2020bvc,
although they do not appear broader.

\subsection{Velocity Estimates from \ion{Fe}{2} Features}
\label{sec:vel}

For each spectrum after $\Delta t=5\,\days$,
we used publicly available code\footnote{https://github.com/nyusngroup/SESNspectraLib} from \citet{Modjaz2016} to measure
the absorption (blueshift) velocities of the
blended \ion{Fe}{2} features at $\lambda\lambda$4924,5018,5169,
which are a proxy for photospheric velocity.
The resulting velocities are listed in Table~\ref{tab:SN2020bvc-spectra}.
Note that the fit did not converge for the NOT spectrum on Mar 02.14, and that we were unable to obtain satisfactory fits for the SEDM spectra.

\begin{figure}[hbt!]
    \centering
    \includegraphics[width=\linewidth]{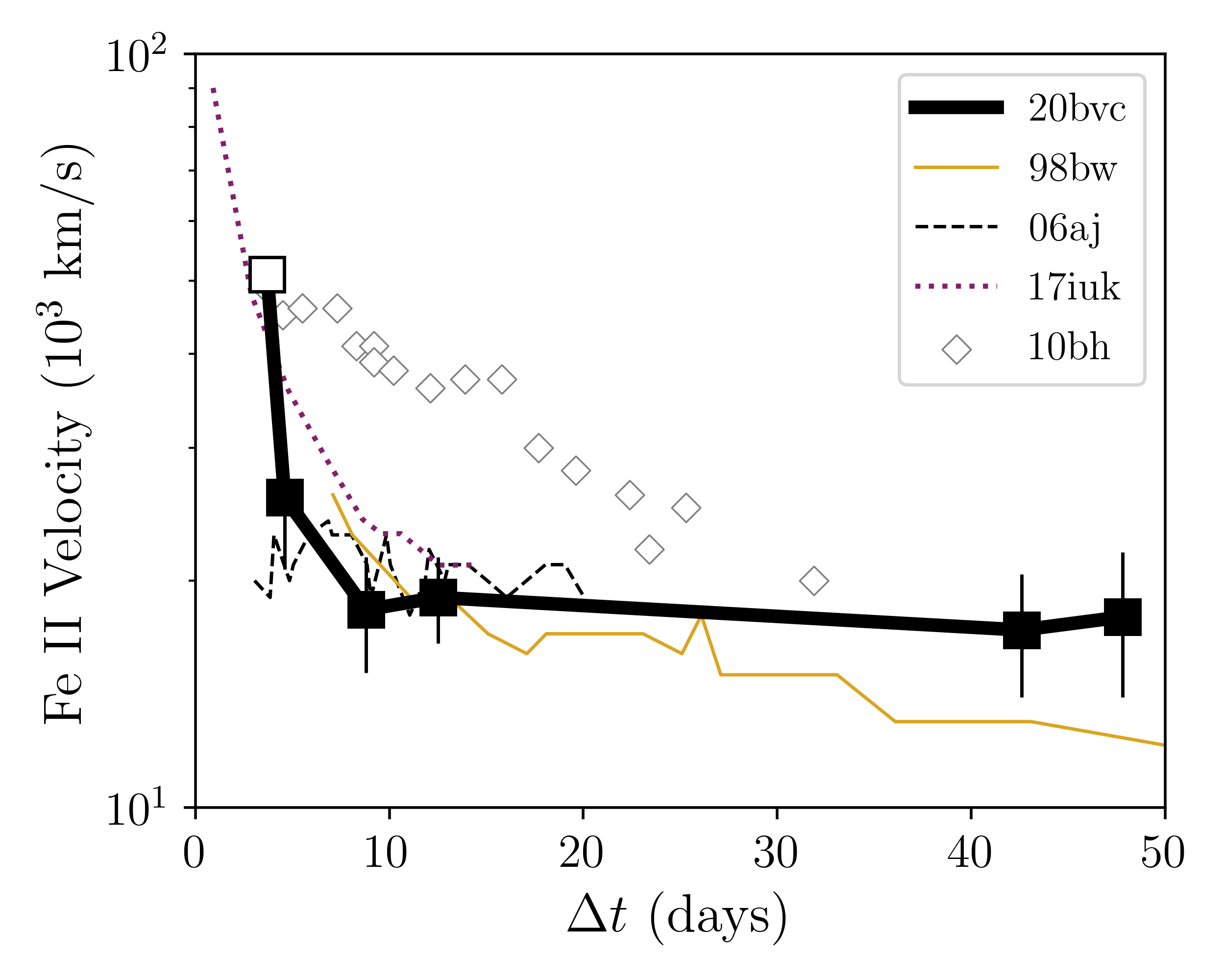}
    \caption{Velocity of SN\,2020bvc (black) compared to LLGRB-SNe. Open symbol corresponds to \ion{Ca}{2} velocity measured from absorption-line minimum,
    and closed symbols correspond to velocities measured by fitting the \ion{Fe}{2} absorption complex.
    Velocities come from \citet{Izzo2019} for SN\,2017iuk and \citet{Modjaz2016} for all other SNe.
    \citet{Modjaz2016} reports velocities from the peak of the optical light curve, so we shifted to time since GRB using \citet{Galama1998} for SN\,1998bw, \citet{Campana2006} for SN\,2006aj, and \citet{Bufano2012} for SN\,2010bh.}
    \label{fig:velocity}
\end{figure}

In Figure~\ref{fig:velocity} we compare the velocity evolution of SN\,2020bvc to that of nearby LLGRB-SNe.
Only SN\,2017iuk and SN\,2020bvc have spectral velocity estimates at early times, and both exhibit a steep drop during the transition from the first to the second optical peak.
During the second peak, the velocities of all but SN\,2010bh are similar to the velocities of Ic-BL SNe associated with GRBs, which are systematically higher than the velocities of Ic-BL SNe lacking associated GRBs \citep{Modjaz2016}.

\section{Modeling the Light Curve}
\label{sec:modeling-opt-lc}

Double-peaked optical light curves have been observed in all types of stripped-envelope SNe:
Type Ic-BL (with SN\,2006aj as the prime example),
Type Ic \citep{Taddia2016,De2018},
Type Ib \citep{Mazzali2008,Chevalier2008,Modjaz2009},
and Type IIb \citep{Arcavi2011,Bersten2012,Bersten2018,Fremling2019}.
The leading explanation for double-peaked light curves in these systems is that the progenitor has as non-standard structure,
with a compact core of mass $M_c$ and low-mass material with $M_e \ll M_c$ extending out to a large radius $R_e$ \citep{Bersten2012,Nakar2014,Piro2015},
although \citet{Sapir2017} have argued that a non-standard envelope structure is not required.

After core-collapse,
a shockwave runs through the thin outer layer,
and in its wake the layer cools (the ``post-shock cooling'' or ``cooling-envelope'' phase),
producing a short-duration first peak.
The remnant is heated from within by the radioactive decay of \nickel\ to \cobalt, which dominates the light curve after a few days, producing the second peak.

In Type~IIb SNe, the extended material is thought to be the stellar envelope.
By contrast, Type~Ic-BL SNe such as SN\,2006aj and SN\,2020bvc are thought to arise from compact stars,
so the envelope is more likely to be extended material that was ejected in a mass-loss episode \citep{Smith2014}.
It is unknown why Ic-BL progenitors would undergo late-stage eruptive mass-loss;
possibilities include binary interaction \citep{Chevalier2012} and gravity waves excited by late-stage convection in the core \citep{Quataert2012}.

Motivated by the similarity between SN\,2006aj and SN\,2020bvc,
we assume that the light curve of SN\,2020bvc is also powered by these two components, and calculate the properties of the explosion and the extended material.

\subsection{Nickel Decay}
\label{sec:nickel}

\begin{figure}[htb!]
    \centering
    \includegraphics[width=\linewidth]{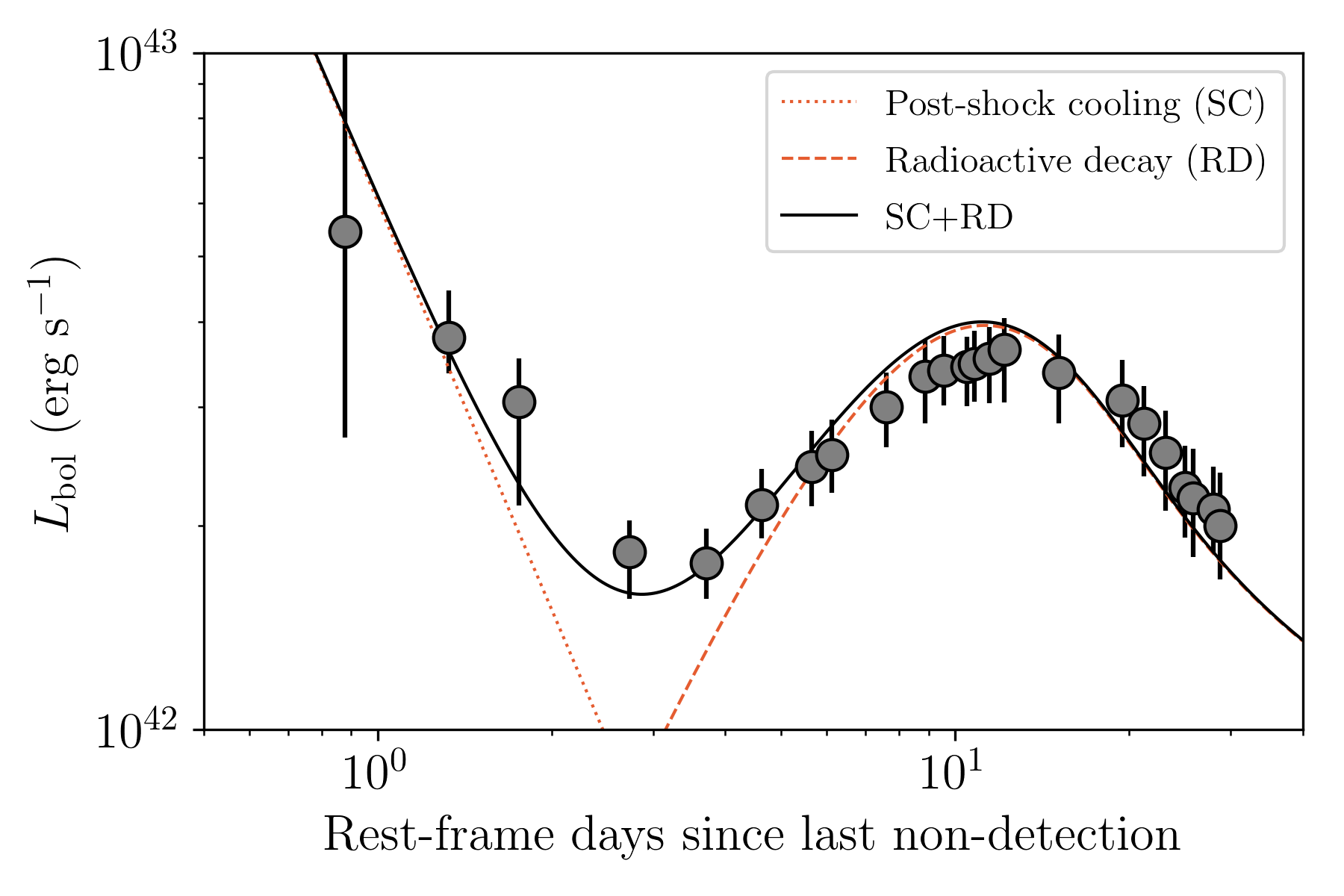}
    \caption{Bolometric luminosity evolution of SN\,2020bvc. The shock-cooling model from \S\ref{sec:shock-cooling} is shown as a dotted orange line. The radioactive decay model from \S\ref{sec:nickel} is shown as a dashed line. The black line is the sum of the two models.}
    \label{fig:bol-lc}
\end{figure}

We use the luminosity and width of the second peak of the SN\,2020bvc light curve to estimate the nickel mass \mni\ and the ejecta mass \mej,
by fitting an Arnett model \citep{Arnett1982}.
Building on the Arnett model,
\citet{Valenti2008} give an analytic formula for $L_\mathrm{bol}(t)$ as a function of \mni\ and a width parameter $\tau_m$, which assumes complete trapping of gamma-rays (not significant in the regime we deal with here).
Fitting the \citet{Valenti2008} light curve to the bolometric light curve from \ref{sec:lc-comparisons}, we obtain $\mni=0.13\pm0.01$ and $\tau_m=8.9\pm0.4$.
The fit is shown in Figure~\ref{fig:bol-lc}.

The value of \mni\ we obtain for SN\,2020bvc
is similar to literature estimates for SN\,2006aj
($\mni=0.20\pm0.10\,\msol$; \citealt{Cano2017}) and
smaller than the nickel mass of SN\,1998bw (0.3--0.6\,\msol; \citealt{Cano2017}),
which is consistent with the relative luminosity of the bolometric light curves (Figure~\ref{fig:bb-evolution}).

Next, we solve for \mej\ and the explosion energy $E_k$ using Equations (2) and (3) in \citet{Lyman2016}.
Taking the opacity $\kappa=0.1\,\cmsq\,\pgram$ (close to the value found from spectral modeling of Ic-BL SNe near peak; \citealt{Mazzali2000}) and $v_\mathrm{ph}=18,000\,\km\,\psec$,
we find $\mej = 2.2 \pm 0.4\,\msol$,
where the uncertainty is dominated by the 20\% uncertainty on $v_\mathrm{ph}$.
The resulting kinetic energy is $E_K = 0.5\,\mej\,v_\mathrm{ph}^2 = 7.1 \pm 2.8 \times 10^{51}\,\erg$.
The explosion parameters for SN\,2020bvc are summarized in Table~\ref{tab:model-summary}.

\begin{deluxetable}{lr}[htb!]
\tablecaption{Explosion properties of SN\,2020bvc \label{tab:model-summary}} 
\tablewidth{0pt} 
\tablehead{\colhead{Parameter} & \colhead{Value}}
\tabletypesize{\normalsize} 
\startdata 
$E_k$ ($10^{51}$\,erg) & $7.1\pm2.8$ \\
$\mej$ (\msol) & $2.2\pm0.4$ \\
$\mni$ (\msol) & $0.13\pm0.01$ \\
$M_e$ (\msol) & $<0.01$ \\
$R_e$ (cm) & $>10^{12}$ \\
\enddata 
\end{deluxetable}

\subsection{Shock cooling}
\label{sec:shock-cooling}

The mass $M_e$ and radius $R_e$ of the material surrounding the progenitor can be estimated using the timescale and luminosity of the first peak.
In \S\ref{sec:spec-evol-comp} we measured a lower limit on the peak bolometric luminosity $L_\mathrm{bol} > 5.62 \times 10^{42}\,\erg\,\psec$,
with an upper limit on the time to peak of 0.7\,d.
From our calculation in Appendix~\ref{sec:appendix-shock-cooling},
we obtain an upper limit on $M_e < 10^{-2}\,\msol$ and a lower limit on $R_e > 10^{12}\cm$.
In Figure~\ref{fig:bol-lc} we show that the bolometric light curve is well described by the sum of the shock cooling model from Appendix~\ref{sec:appendix-shock-cooling} with $R_e = 4 \times 10^{12}\,\cm$ and $M_e = 10^{-2}\,\msol$,
and a \nickel-powered light curve with the properties calculated in \S\ref{sec:nickel}.
The shock-cooling light curve only describes the decline after peak;
we do not attempt to model the rise.
The properties of the ambient material
are summarized in Table~\ref{tab:model-summary}.

The values of $M_e$ and $R_e$ we measured for SN\,2020bvc
are consistent with what was inferred for SN\,2006aj, which had much more detailed early UV and optical data: $M_e=4\times10^{-3}\,\msol$ and $R_e = 9 \times 10^{12}\,\cm$ \citep{Irwin2016}.
A similar low-mass shell was inferred for the Ic-BL SN\,2018gep \citep{Ho2019gep}:
in that case, the shell ($M_e = 0.02\,\msol$) was at a larger radius ($R_e=3\times10^{14}\,\cm$), which prolonged the shock-interaction peak and blended it with the \nickel-powered peak.
A similarly low-mass, large-radius shell may also explain the
luminous light curve of the Ic-BL SN iPTF16asu \citep{Whitesides2017}.
With these four events, we may be seeing a continuum in shell properties around Ic-BL SNe,
resulting from different mass-loss behavior shortly prior to core-collapse \citep{Smith2014}.

\section{Modeling the Fast Ejecta}
\label{sec:modeling-radio}

One of the key features of LLGRB\,060218 / SN\,2006aj
was radio and X-ray emission that peaked earlier and was more luminous than that of ordinary CC SNe.
Here we compare the early (1--50\,d) radio and X-ray properties of SN\,2020bvc to that of SN\,2006aj and other LLGRB-SNe.

\subsection{Radio Emission}

We have several reasons to believe that the radio emission is dominated by the transient rather than by the host galaxy.
First, the flux density is observed to decline at 6\,\ghz\ and 10\,\ghz, albeit marginally.
Second, in \S\ref{sec:obs} we found that the source is unresolved (i.e. a point source) at all frequencies.
Third, at all frequencies
the centroid of the radio source is consistent with the position of the optical transient, and there is no other radio source detected in the vicinity of the galaxy.
(There is a nearby \ion{H}{2} region, but this would produce free-free emission and therefore a flat spectral index, which is inconsistent with our observations.)
Late-time radio observations will be used to be secure, and to subtract any host contribution.

If the emission at 3\,\ghz\ were entirely from the underlying host-galaxy region (the synthesized beamwidth at this frequency is 7\arcsec)
the flux density at this frequency can be used to estimate a star-formation rate of 0.2\,\msol\,\pyr\ using the prescription in \citep{Greiner2016,Murphy2011}:

\begin{equation}
\begin{split}
    \left( \frac{\mathrm{SFR}_\mathrm{Radio}}{\msol\,\pyr} \right) 
    & = 0.059
    \left( \frac{F_\nu}{\mathrm{\mu Jy}} \right) (1+z)^{-(\alpha+1)} \\
    & \times \left( \frac{D_L}{\mathrm{Gpc}} \right)^2 \left( \frac{\nu}{\ghz} \right)^{-\alpha}
\end{split}
\end{equation}

\noindent where we use $F_\nu=120\,\mu$Jy, $\nu=3\,\ghz$, and $\alpha=-0.9$ for $F_\nu \propto \nu^{\alpha}$.

\begin{figure}[hbt!]
    \centering
    \includegraphics[width=\linewidth]{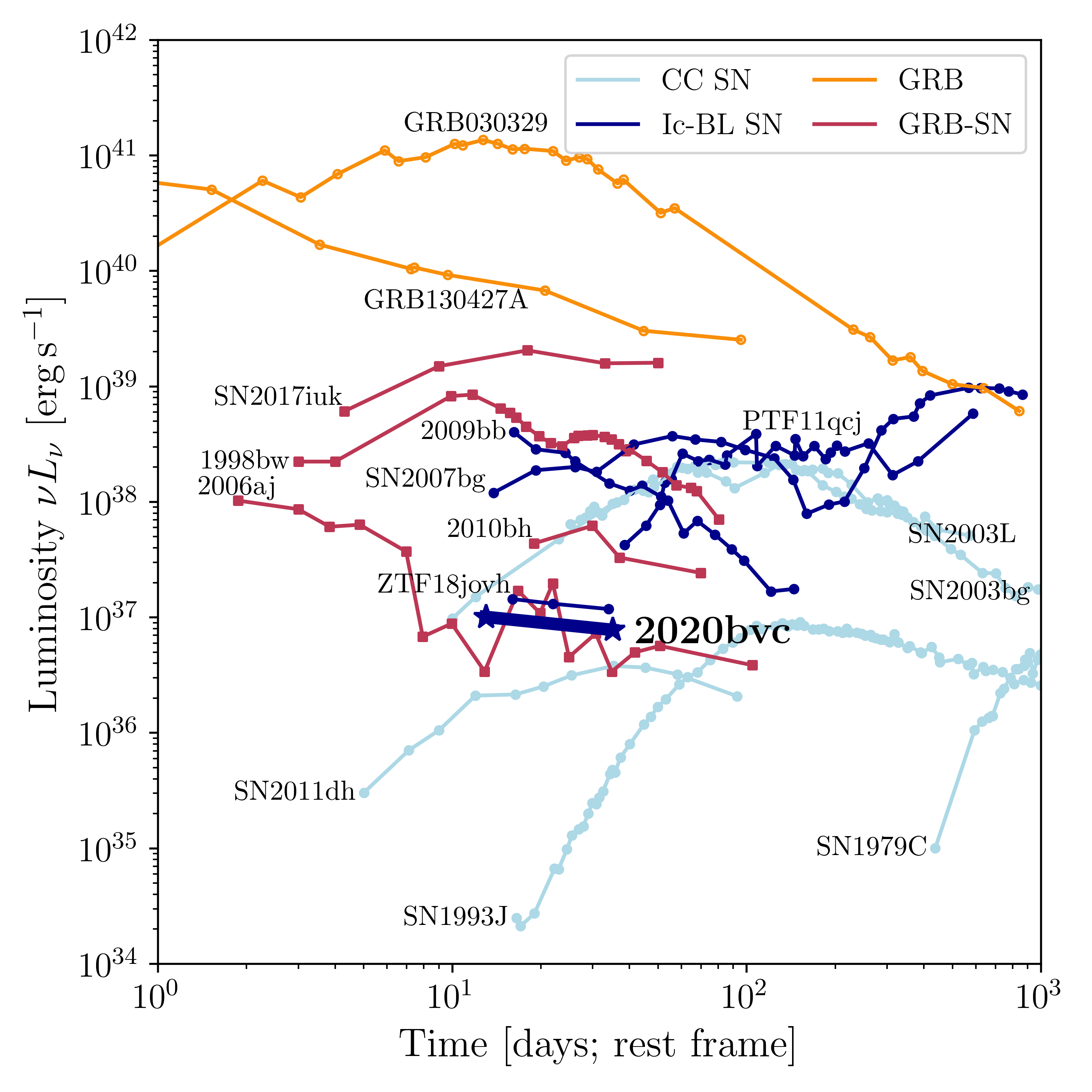}
    \caption{10\,\ghz\ radio light curve of SN\,2020bvc (points) compared to low-luminosity GRBs and relativistic Ic-BL SNe.
    Light curve of GRB\,130427A is the 6.8\,\ghz\ light curve from \citet{Perley2014}.
    Data point for SN\,2017iuk is at 6\,\ghz\ \citep{Laskar2017}.
    SN\,2006aj data is at 8.5\,\ghz\ from \citet{Soderberg2006aj}.
    ZTF18aaqjovh data is from \citet{Ho2020jovh}.
    SN\,2010bh light curve is at 5.4\,GHz from \citet{Margutti2014}.
    PTF\,11qcj light curve is at 5\,GHz from \citet{Corsi2014}.
    All other sources are as described in Appendix C of \citet{Ho2019cow}.
    }
    \label{fig:radio-lc}
\end{figure}

For now, we assume that the radio emission is primarily from the transient.
In Figure~\ref{fig:radio-lc} we show the 10\,\ghz\ radio light curve of SN\,2020bvc.
The luminosity is similar to that of SN\,2006aj and SN\,2010bh,
and significantly fainter than that of SN\,2017iuk, SN\,1998bw.
In \citet{Ho2019cow} we found that the radio luminosity is directly proportional to $U/R$, the (thermalized) energy of the blastwave divided by the shock radius.
So, the lower radio luminosity of SN\,2006aj and SN\,2020bvc could correspond to a lower explosion energy.
This is consistent with the finding in
\S\ref{sec:nickel} that SN\,2006aj and SN\,2020bvc have a similar kinetic energy, which is significantly smaller than the kinetic energy of SN\,1998bw and SN\,2017iuk.

From the radio SED, we estimate that the
peak frequency is $<3\,\ghz$ at $\Delta t=24\,\days$, with a peak flux density $>113\,\mu$Jy.
We use these values and the framework described in \citet{Chevalier1998} to estimate properties of the forward shock and ambient medium.
We list the results in Table~\ref{tab:radio-xray-model-summary},
discuss the implications here,
and provide the calculation in Appendix~\ref{sec:appendix-radio-modeling}.
In Figure~\ref{fig:lum-tnu} we show the peak frequency and time compared to the peak luminosity,
with lines indicating how these values correspond to ambient density (mass-loss rate) and energy.

\begin{figure}[hbt!]
    \centering
    \includegraphics[width=\linewidth]{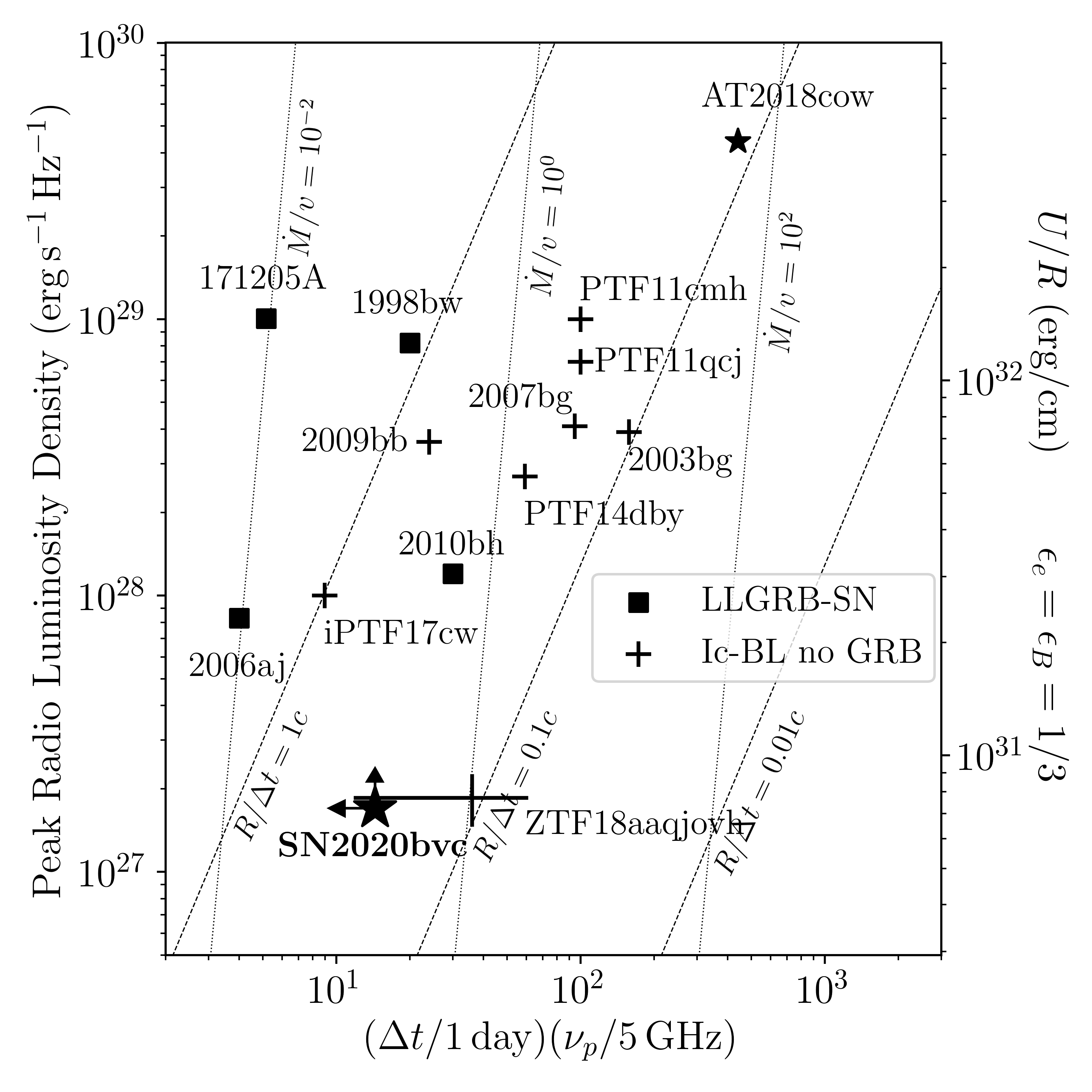}
    \caption{Luminosity and peak frequency of the radio light curve of SN\,2020bvc compared to LLGRBs and energetic SNe.
    Lines of constant mass-loss rate (scaled to wind velocity) are shown in units of $10^{-4}\,\msol\,\pyr/1000\,\km\,\psec$.
    Data for PTF14dby are from 7.4\,\ghz\ light curve in \citet{Corsi2016}.
    Data for PTF11cmh and PTF11qcj are from 5\,\ghz\ light curve in \citet{Corsi2016}.
    Data for iPTF17cw are from the 2.8\,\ghz\ light curve in \citet{Corsi2017}.
    Data for ZTF18aaqjovh are from \citet{Ho2020jovh}.
    For details on all other sources,
    see caption to Figure~5 and Appendix~C in \citet{Ho2019cow}.
    }
    \label{fig:lum-tnu}
\end{figure}

\begin{deluxetable}{lrr}[htb!]
\tablecaption{Properties of the forward shock in SN\,2020bvc derived from radio and X-ray observations at $\Delta t=24\,\days$  \label{tab:radio-xray-model-summary}} 
\tablewidth{0pt} 
\tablehead{\colhead{Parameter} & \colhead{Value}}
\tabletypesize{\normalsize} 
\startdata 
$\nu_a = \nu_p$ (GHz) & $<3$ \\
$F_{\nu,p}$ ($\mu\jy$) & $>110$ \\
$R$ $(\cm)$ & $>1.7 \times 10^{16}$ \\
$v/c$ & $>0.3$ \\
$B$ (G) & $<0.34$ \\
$U$ ($\erg$) & $1.5 \times 10^{47}$ \\
$n_e$ ($\pcmcub$) & $160$ \\
$\nu_c$ (Hz) & $1.4 \times 10^{13}$ \\
\enddata 
\end{deluxetable}

First, we find a forward shock radius of $1.7 \times 10^{16}\,\cm$,
implying
a mean velocity up to 24\,\days\ of 
$\Gamma \beta > 0.28$.
As shown in Figure~\ref{fig:lum-tnu},
the lower limit on the velocity we infer is similar to the mildly relativistic velocities inferred for some LLGRB-SNe, in particular SN\,2010bh.
It is also possible that the velocity approaches the
relativistic speeds inferred for SN\,2006aj and SN\,1998bw.

Second, we find a lower limit on the energy thermalized by the shock of $1.3 \times 10^{47}\,\erg$. As shown in Figure~\ref{fig:lum-tnu}, SN\,2020bvc appears to have an energy most similar to that of SN\,2006aj and a radio-loud Ic-BL SN recently discovered in ZTF \citep{Ho2020jovh}.

Third, we find an ambient density of 
$n_e=160\,\pcmcub$, which we show in Figure~\ref{fig:lum-tnu} as a mass-loss rate of $\sim10^{-5}\,\msol\,\pyr$, assuming a wind velocity $v_w=1000\,\km\,\psec$.
As shown in the figure, this mass-loss rate is within an order of magnitude of LLGRB-SNe, including SN\,2006aj, SN\,1998bw, and SN\,2010bh.

Fourth, we find that the cooling frequency is $\nu_c = 1.0 \times 10^{13}\,$Hz, below the X-ray band.
We discuss the implications in \S\ref{sec:xray}.

Finally, we address the model proposed in \citet{Izzo2020}, that SN\,2020bvc represents a GRB jet with energy $2 \times 10^{51}\,\erg$ viewed at an angle of 23 degrees ($\theta_\mathrm{obs}=0.4$), propagating into a power-law density profile $R^{-1.5}$.
The authors argue that this event has similar early optical behavior to LLGRB\,171205A / SN\,2017iuk and that the X-ray emission is consistent with the predicted light curve from \citet{Granot2018}.
We point out that the same model predicts an 8.5\,\ghz\ radio light curve that exceeds $10^{30}\,\erg\,\psec\,\phz$ over the period of our VLA observations, several orders of magnitude more luminous than our measurements.
An off-axis jet cannot be entirely ruled out; future radio observations will be needed to determine whether a highly off-axis jet could be present.
However, for now we find that no off-axis jet is required to explain the 1--50\,d radio light curve, as was the case for SN\,2006aj \citep{Soderberg2006aj}.
To our knowledge only one radio data point has been published for SN\,2017iuk,
and the radio emission compared to off-axis models was not discussed in \citet{Izzo2019}.

In conclusion, the radio properties of SN\,2020bvc are similar to what has been observed for LLGRB-SNe. Although we do not have evidence for relativistic ejecta or a GRB, the radio light curve is unlike what has been seen for ``ordinary'' core-collapse SNe, suggesting that SN\,2020bvc is related to the LLGRB phenomenon,
i.e. an LLGRB-like event discovered optically.

\subsection{X-ray Emission}
\label{sec:modeling-xray}

In this section we compare the X-ray light curve,
and the X-ray to radio SED,
of SN\,2020bvc to that of SN\,2006aj and other LLGRBs in the literature.

\begin{figure}
    \centering
    \includegraphics[width=\linewidth]{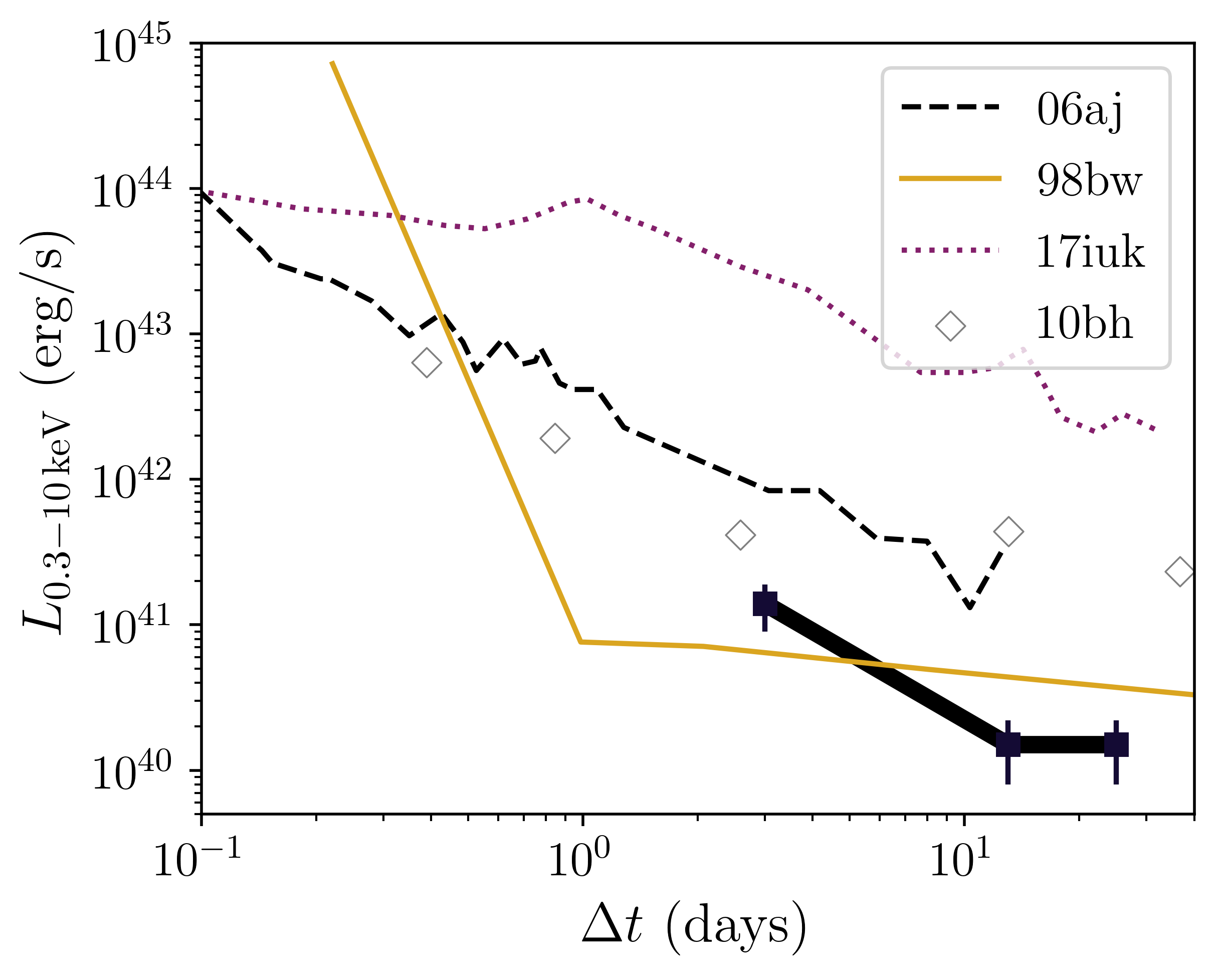}
    \caption{The 0.3--10\,keV X-ray light curve of SN\,2020bvc (black connected squares) compared to that of nearby Ic-BL SNe associated with LLGRBs. Data on GRB-SNe taken from \citet{Campana2006}, \citet{Corsi2017}, and \citet{Delia2018}.}
    \label{fig:xray-comparison}
\end{figure}

The X-ray light curve of LLGRB\,060218 / SN\,2006aj had two components:
the prompt emission itself, which lasted until $10^{4}\,$s (often called a GRB, but given the low peak energy is also called an X-Ray Flash or XRF) and an afterglow that decayed as $t^{-\alpha}$ where $\alpha=1.2\pm0.1$ until $10^{6}\,$s \citep{Campana2006,Soderberg2006aj}.
The 0.3--10\,\kev\ luminosity was $8 \times 10^{41}\,\erg\,\psec$ at three days post-explosion \citep{Campana2006}.
In Figure~\ref{fig:xray-comparison} we show the 0.3--10\,\kev\ light curve of SN\,2020bvc compared to that of SN\,2006aj and nearby LLGRB-SNe.
We find that the X-ray luminosity is within an order of magnitude of SN\,2006aj, as well as SN\,1998bw and SN\,2010bh.

Next we consider the radio to X-ray spectral index.
At $\Delta t=13\,\days$ the radio to X-ray spectral index of SN\,2020bvc is $\beta_{RX}=0.5$,
where $F_\nu \propto \nu^{-\beta}$.
Given that the cooling frequency lies below the X-ray band (\S\ref{sec:modeling-radio}) the value of  $\beta_{RX}$ is too shallow for the X-rays to be an extension of the radio synchrotron spectrum.
The same was true of SN\,2006aj, which had a very similar value of $\beta_{RX}=0.5$ \citep{Soderberg2006aj,Fan2006,Irwin2016}.

\begin{figure}
    \centering
    \includegraphics[width=\linewidth]{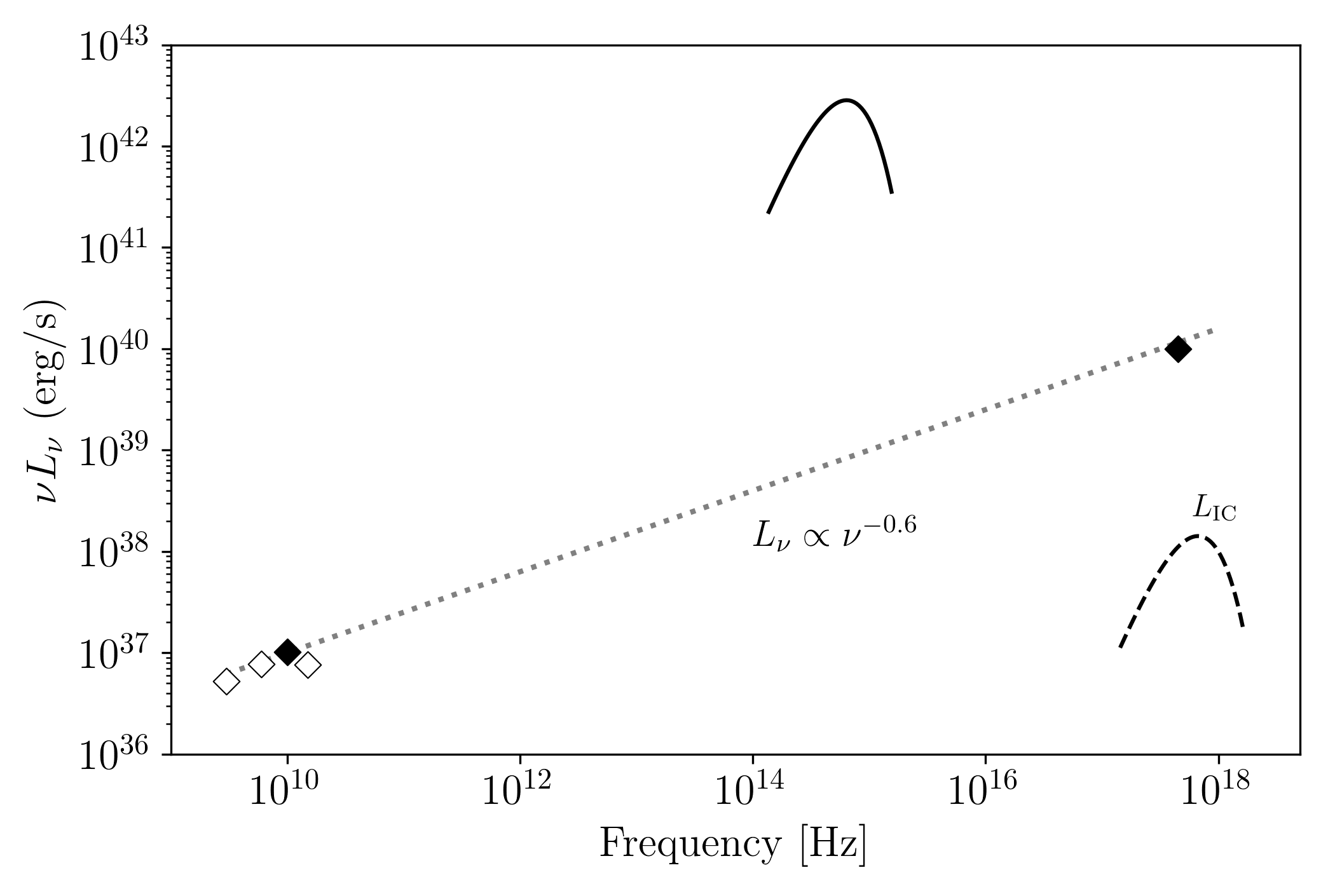}
    \caption{The SED from radio to X-rays at $\Delta t=13\,\days$.
    The empty diamonds are VLA data points from 17--28\,\days.
    The solid line is the blackbody fit to the optical SED.
    The dotted line shows an extrapolation of $L_\nu \propto \nu^{-(p-1)/2}$ where $p=2.2$,
    and the dashed curve shows the predicted emission from inverse Compton scattering (calculated in Appendix~\ref{sec:appendix-ic}).}
    \label{fig:sed}
\end{figure}

Furthermore, for the X-rays to be an extension of the synchrotron spectrum we would require
$\nu_c > 10^{17}\,\hz$ at $t\approx30\,\days$
and therefore $B<0.01\,\gauss$,
which is over an order of magnitude smaller than the value of $B$ measured in any known SN \citep{Chevalier1998,Chevalier2006,Corsi2016}.
This is another argument for why the X-rays are unlikely to arise from the same synchrotron spectrum as the radio emission.

Finally, from the ratio of the optical to radio luminosity, we can estimate the expected contribution of X-rays from inverse Compton scattering.
We find (Appendix~\ref{sec:appendix-ic}) that the contribution is not sufficient to explain the X-ray luminosity that we observe, which again was also the case in SN\,2006aj.
The X-ray ``excess'' observed in SN\,2006aj has been
attributed to the long-lived activity of a central engine \citep{Soderberg2006aj,Fan2006} and to dust scattering \citep{Margutti2015,Irwin2016}.
On the other hand, \citet{Waxman2007} argued that the long-lived X-ray emission could be explained naturally in a model of mildly relativistic shock breakout into a wind, and that it was the radio emission that required a separate component.
The data we have are less detailed than that obtained for SN\,2006aj, so are not useful in distinguishing between these different possibilities.

\section{Early ZTF Light Curves of Nearby Ic-BL SNe}
\label{sec:early-ztf-lc}

As discussed in \S\ref{sec:shock-cooling}, the timescale and luminosity of the shock-cooling peak
is most sensitive to the shell properties (mass, radius) and the shock velocity.
By contrast, the timescale and luminosity of the radioactively-powered peak is set by the nickel mass,
the ejecta mass, and the explosion energy.
So, it is not obvious that the properties of the second peak (which are heterogeneous;  \citealt{Taddia2019}) should be correlated with the properties of the first peak.

\begin{figure*}[hbt!]
    \centering
    \includegraphics[width=0.8\textwidth]{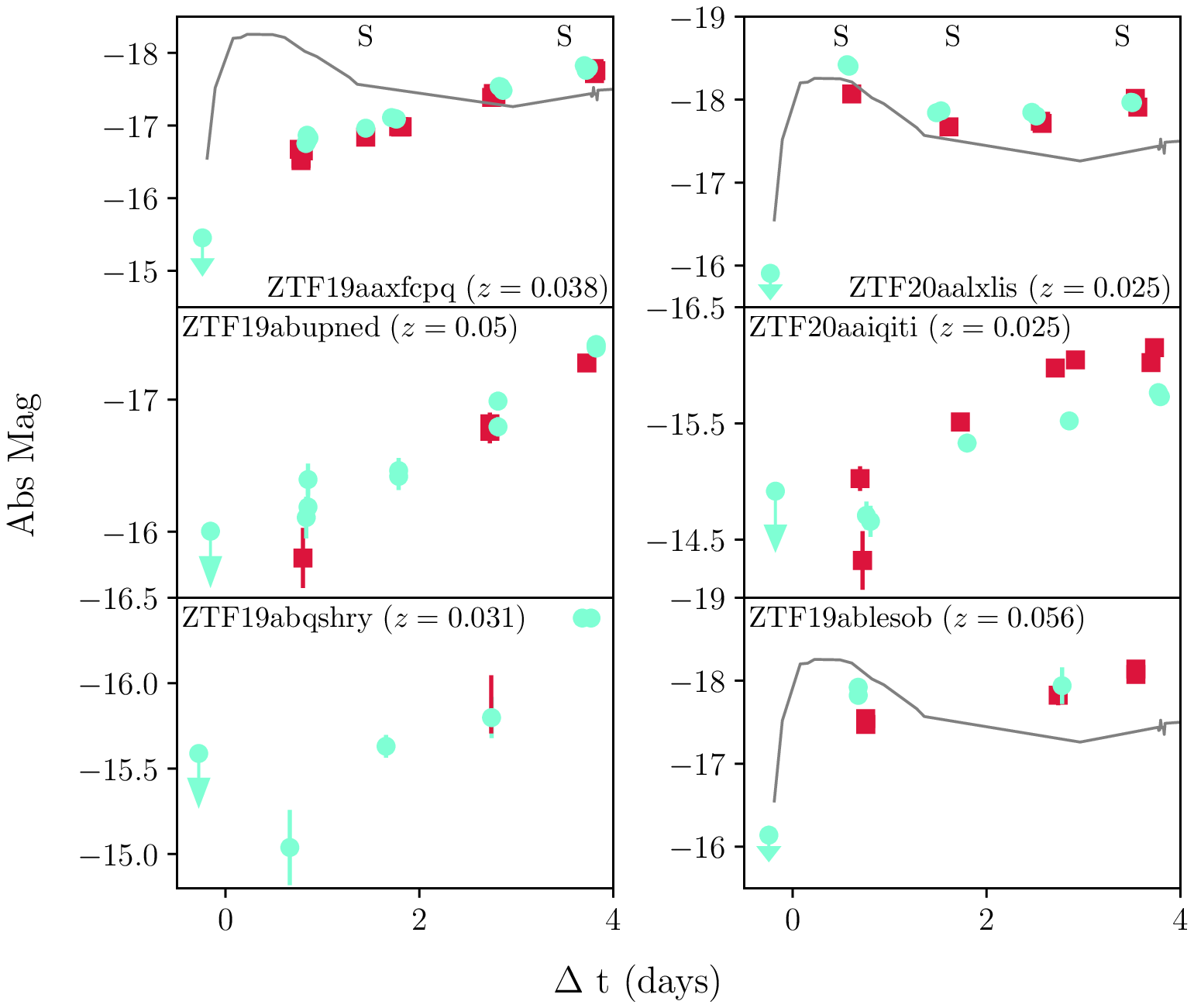}
    \caption{Early ($\Delta t \lesssim 4\,\days$) light curves of nearby Ic-BL SNe observed as part of ZTF's high-cadence surveys, from forced photometry on P48 images \citep{Yao2019}. The $B$-band light curve of SN\,2006aj is shown as a grey line for comparison. Epochs of follow-up spectroscopy are marked with `S' along the top of the panel.}
    \label{fig:ztf-lc}
\end{figure*}

In Figure~\ref{fig:ztf-lc} we show early ($<4\,\days$) light curves of five nearby ($z\lesssim0.05$) Ic-BL SNe observed as part of ZTF's high-cadence surveys,
which were spectroscopically classified as part of the ZTF flux-limited \citep{Fremling2019rcf} and volume-limited \citep{De2020_clu} experiments.
The light curves shown are
from forced photometry on P48 images \citep{Yao2019}, and epochs of spectroscopy are marked with `S.'
For the two most luminous events, we show the light curve of SN\,2006aj for comparison.
We can rule out a first peak like that of SN\,2006aj (duration $\approx 1\,\days$, peak luminosity $\approx -18$) for all events except 
one (ZTF19ablesob).
Note that the faintest LLGRB-SN, SN\,2010bh, peaked at $M=-17\,$mag:
with the ZTF flux-limited survey we would be over 90\% complete for such events out to $z=0.03$.
SN\,2020bvc peaked brighter than $M=-18.5$,
so the flux-limited survey would be over 90\% complete for such events out to $z=0.06$.

Our high-cadence optical observations provide the first evidence that Ic-BL SNe like LLGRB\,060218/SN\,2006aj are not the norm.
Radio follow-up observations have only been sensitive enough to show that events like LLGRB\,980425/SN\,1998bw are uncommon \citep{Corsi2016}, and in most cases have been unable to rule out emission like that seen in SN\,2006aj and SN\,2020bvc.

There are many degeneracies that complicate the interpretation of Figure~\ref{fig:ztf-lc}.
Rise time and peak luminosity are sensitive to the velocity of the shock.
The shock velocity when it breaks out of the star is sensitive to the outer density gradient in the stellar envelope and the energy of the explosion.
Even if all Ic-BL progenitors were identical,
there could be a strong dependence with observing angle.
Ic-BL SNe are expected to be asymmetric and bipolar,
so the ejecta directed along the poles will move faster than along the equator.
Thus, an event viewed along the poles could have a much brighter shock-interaction peak.

Finally, assuming identical and spherically symmetric explosions for the Ic-BL SNe,
there could be wide diversity in properties of the ambient material, i.e. mass, radius, and geometry.
The circumstellar medium (CSM) itself could be asymmetric (e.g. a disk rather than a spherical wind) introducing even more complicated viewing-angle effects.

As we discussed in our analysis of another fast-rising luminous Ic-BL SN, SN\,2018gep \citep{Ho2019gep},
it can be difficult to know when it is appropriate to model such emission as arising from shock breakout in low-mass, large-radius material,
and when it is appropriate to model such emission as arising from post-shock cooling in higher-mass, smaller-radius material (e.g. \citealt{Nakar2014,Piro2015}).
In short, it is extremely difficult at present to explain why we see such diversity in the early light curves of Ic-BL SNe.
A model grid of different explosion and CSM properties, with resulting light curves, is in preparation (Khatami et al. in prep) and will be useful in understanding what configurations are ruled out or allowed for each of the objects in Figure~\ref{fig:ztf-lc}.

\section{Summary and Discussion}
\label{sec:summary}

We presented optical, X-ray, and radio observations of SN\,2020bvc, which shares key characteristics with the Ic-BL SN\,2006aj associated with LLGRB\,060218.
Both events had:

\begin{itemize}

    \item A double-peaked optical light curve. The first peak is fast ($\approx 1\,\days$), luminous ($M=-18$), and blue ($g-r\approx-0.3\,$mag),
    and can be modeled as shock-cooling emission
    from low-mass ($M_e < 10^{-2}\,\msol$) extended ($R_e > 10^{12}\,\cm$) material;
    
    \item Radio emission ($10^{37}\,\erg\,\psec$ at 10\,\ghz) from a mildly relativistic ($v > 0.3c$) forward shock, much fainter than that observed in LLGRB-SNe such as SN\,1998bw and SN\,2017iuk; and
    
    \item X-ray emission of a similar luminosity ($10^{41}\,\erg\,\psec$) that likely requires a separate emission component from that producing the radio emission.
    
\end{itemize}

When our paper was nearly complete, \citet{Izzo2020} presented an interpretation of SN\,2020bvc as a classical high-energy ($2\times10^{51}\,$erg) GRB viewed 23\,degrees off-axis on the basis of (1) the fast expansion velocities ($v_\mathrm{exp}\approx70,000\,\km\,\psec$) measured from the early optical spectra, similar to those observed in the Ic-BL SN\,2017iuk accompanying LLGRB\,171205A,
(2) the X-ray light curve,
and (3) the double-peaked UVOT light curve, where the first peak was argued to arise from
the cocoon expanding and cooling after breaking out of the progenitor star.
In our work we found that from the perspective of the radio observations obtained so far (1--50\,d post-discovery),
no off-axis jet is required.
In particular, the faint radio light curve is not consistent with the model in \citet{Granot2018} invoked by \citet{Izzo2020} to explain the X-ray data.

Instead, the simplest explanation from our data is that SN\,2020bvc is a similar event to LLGRB\,060218/SN\,2006aj.
LLGRB\,060218/SN\,2006aj has been extensively modeled
and a summary of leading interpretations can be found in \citet{Irwin2016}.
Here we outline the different models,
then discuss how high-cadence optical surveys, together with early spectroscopy and X-ray and radio follow-up observations, can help distinguish between them.

\begin{enumerate}[label=(\alph*)]

    \item \emph{Mildly relativistic shock breakout into a wind.} \citet{Campana2006} and \citet{Waxman2007} proposed that this single mechanism was responsible for the LLGRB, the shock-cooling emission, and the X-ray afterglow, in which case all three would be isotropic (a different low-energy component would be needed for the radio emission).
    
    \item \emph{Choked jet.} \citet{Nakar2015} expanded on the model above by suggesting that the shock breakout is powered by an energetic GRB-like jet that is choked in extended low-mass material surrounding the progenitor star. Again, all emission components would be expected to be isotropic.
    
    \item \emph{On-axis low-power jet.} \citet{Irwin2016} proposed that the LLGRB and the shock-cooling emission are decoupled:
    the LLGRB was produced by a successful collimated low-power jet, and the shock-cooling emission by spherical SN ejecta. In that case, the LLGRB would only be observable within a small viewing angle,
    while the shock-cooling emission would be isotropic.
    
\end{enumerate}

In \S\ref{sec:early-ztf-lc} we found that a number of Ic-BL lack luminous early peaks.
If X-ray and radio observations of such events reveal LLGRB-like X-ray and radio emission,
and the shock-cooling emission is indeed expected to be isotropic,
this would argue against a single mechanism for the shock-cooling emission and the afterglow.
If, on the other hand, a double-peaked optical light curve is predictive of LLGRB-like X-ray and radio emission, and single-peaked events lack such emission, that would support models in which these components are produced by the same mechanism.
Another test is the relative rates:
if the LLGRB is only observable within a small viewing angle, the rate of double-peaked Ic-BL SNe should significantly exceed the rate of LLGRBs.

The key argument that LLGRB\,171205A / SN\,2017iuk arose from a jet was the presence of iron-peak elements in the early spectra, thought to have been transported to the surface by the jet \citep{Izzo2019}.
SN\,2017iuk was discovered via a GRB trigger,
but with SN\,2020bvc we have demonstrated that high-cadence optical surveys can enable similarly early spectroscopic observations.
So, it should be possible to search for these cocoon signatures for a larger sample of events,
without relying on the detection of an LLGRB.
For events with detected cocoon emission,
the long-term radio light curve is crucial for distinguishing between off-axis jets and choked jets. 

We point out that based on estimated rates of GRBs and LLGRBs,
the rate of off-axis GRBs in the local universe ($z<0.05$) is only one order of magnitude smaller than the rate of LLGRBs \citep{Soderberg2006aj, Liang2007}, which are detected routinely (if infrequently--see the discussion below regarding why).
The estimated rate of on-axis GRBs at $z=0$ is $0.42^{+0.90}_{-0.40}\,\pyr\,\pgpccub$,
as measured from the \swift\ sample of classical GRBs \citep{Lien2014}.
Taking a beaming fraction of 0.01 \citep{Guetta2005} the expectation is for two (and up to six) GRBs in the local universe per year.
Recently, \citet{Law2018} identified a candidate off-axis GRB afterglow in data from the VLA Sky Survey.
Their estimate of the rate of events similar to this off-axis candidate is consistent with the expected off-axis GRB rate in the local universe.

Unfortunately, bursts like LLGRB\,060218 are difficult to detect with ongoing GRB satellites, which are tuned to finding cosmological GRBs.
First, the low luminosity ($L_\mathrm{iso}=2.6\times10^{46}\,\erg\,\psec$) means that an LLGRB like 060218 can only be detected in the nearby universe.
Second, the long timescale ($T_{90}=2100\,$s) makes it difficult to detect the event above the background evolution of wide-field detectors.
Third, the low peak energy ($E_\mathrm{pk}=5\,$keV) means that the burst is at the bottom of the energy range for sensitive wide-field detectors like \fermi/GBM and the Interplanetary Network \citep{Hurley2010}.
Finally, the fact that a burst like 060218 would only be detected in the local universe means that the number $N$ detectable above a flux threshold $S$ goes as $\log(N>S)\propto S^{-3/2}$:
the number detected is very sensitive to the threshold used.
Going forward, it would be useful to have a wide-field mission optimized for the detection of low-luminosity, long-duration bursts that peak in the soft X-ray band.

Due to the low LLGRB discovery rate and the small sample size, the LLGRB rate is highly uncertain;
it is currently roughly consistent with the rate of Ic-BL SNe \citep{Li2011,Kelly2012}.
An outstanding question is therefore whether all Ic-BL SNe harbor an LLGRB.
The effort to answer this question has been led by radio follow-up observations:
by following up dozens of Ic-BL SNe found in wide-field optical surveys,
\citet{Corsi2016} limited the fraction
harboring SN\,1998bw-like radio emission to
$\lesssim 14\%$ \citep{Corsi2016}.
However, as shown in Figure~\ref{fig:radio-lc},
SN\,1998bw was the most radio-luminous LLGRB-SN.
Radio observations have generally not been sensitive enough to rule out a radio counterpart like that accompanying SN\,2006aj.

High-cadence optical surveys provide a novel opportunity to measure the rate of Ic-BL SNe that are similar to SN\,2006aj.
Optical shock-cooling emission is expected to be isotropic,
and should not depend on the explosion properties that determine the second peak (ejecta mass, nickel mass).
From the events in ZTF with early high-cadence light curves, it appears that
SN\,2006aj-like events are uncommon,
but more events will be needed to measure a robust rate.

The code used to produce the results described
in this paper was written in Python and is available
online in an open-source repository\footnote{https://github.com/annayqho/SN2020bvc}.

\appendix

\section{Photometry Table}
\label{sec:appendix-phot-table}

In Table~\ref{tab:uvot-phot} we provide the complete UVOIR photometry for SN\,2020bvc.

\startlongtable 
\begin{deluxetable*}{lrrrr}
\tablecaption{UVOIR photometry for SN\,2020bvc, corrected for Milky Way extinction. Epochs given in observer-frame since $t_0$ (defined in \S\ref{sec:ztf-detection}) \label{tab:uvot-phot}} 
\tablewidth{0pt} 
\tablehead{ \colhead{Date} & \colhead{$\Delta t$} & \colhead{Inst.} & \colhead{Filt.} & \colhead{Mag} \\ \colhead{(MJD)} & \colhead{(d)} & \colhead{} & \colhead{} & \colhead{(AB)}} 
\tabletypesize{\scriptsize} 
\startdata 
58883.3406 & 0.67 & P48+ZTF & $i$ & $17.44 \pm 0.05$ \\ 
58883.3901 & 0.72 & P48+ZTF & $i$ & $17.46 \pm 0.04$ \\ 
58883.4763 & 0.81 & P48+ZTF & $g$ & $16.82 \pm 0.04$ \\ 
58883.4966 & 0.83 & P48+ZTF & $g$ & $16.83 \pm 0.05$ \\ 
58883.524 & 0.85 & P48+ZTF & $r$ & $17.19 \pm 0.04$ \\ 
58884.0245 & 1.35 & Swift+UVOT & $UVW1$ & $17.15 \pm 0.01$ \\ 
58884.0253 & 1.36 & Swift+UVOT & $U$ & $17.08 \pm 0.01$ \\ 
58884.0257 & 1.36 & Swift+UVOT & $B$ & $17.23 \pm 0.01$ \\ 
58884.0268 & 1.36 & Swift+UVOT & $UVW2$ & $17.90 \pm 0.01$ \\ 
58884.028 & 1.36 & Swift+UVOT & $V$ & $17.12 \pm 0.01$ \\ 
58884.0297 & 1.36 & Swift+UVOT & $UVM2$ & $17.39 \pm 0.01$ \\ 
58884.1362 & 1.47 & LT+IOO & $g$ & $17.30 \pm 0.01$ \\ 
58884.3634 & 1.69 & P60+SEDM & $i$ & $17.50 \pm 0.03$ \\ 
58884.3889 & 1.72 & P48+ZTF & $i$ & $17.66 \pm 0.05$ \\ 
58884.4109 & 1.74 & P48+ZTF & $i$ & $17.63 \pm 0.04$ \\ 
58884.4212 & 1.75 & P48+ZTF & $g$ & $17.40 \pm 0.06$ \\ 
58884.469 & 1.8 & P48+ZTF & $g$ & $17.38 \pm 0.05$ \\ 
58884.4754 & 1.81 & P48+ZTF & $g$ & $17.37 \pm 0.05$ \\ 
58884.5473 & 1.88 & P48+ZTF & $r$ & $17.58 \pm 0.06$ \\ 
58884.5533 & 1.88 & P48+ZTF & $r$ & $17.57 \pm 0.04$ \\ 
58885.3891 & 2.72 & P48+ZTF & $i$ & $17.67 \pm 0.06$ \\ 
58885.4111 & 2.74 & P48+ZTF & $i$ & $17.65 \pm 0.04$ \\ 
58885.429 & 2.76 & P48+ZTF & $g$ & $17.40 \pm 0.05$ \\ 
58885.4774 & 2.81 & P48+ZTF & $g$ & $17.44 \pm 0.07$ \\ 
58885.5211 & 2.85 & P48+ZTF & $r$ & $17.51 \pm 0.04$ \\ 
58885.538 & 2.87 & P48+ZTF & $r$ & $17.52 \pm 0.05$ \\ 
58885.5533 & 2.88 & Swift+UVOT & $UVW1$ & $19.48 \pm 0.01$ \\ 
58885.554 & 2.88 & Swift+UVOT & $U$ & $18.33 \pm 0.01$ \\ 
58885.5543 & 2.88 & Swift+UVOT & $B$ & $17.48 \pm 0.01$ \\ 
58885.5553 & 2.89 & Swift+UVOT & $UVW2$ & $20.03 \pm 0.01$ \\ 
58885.5563 & 2.89 & Swift+UVOT & $V$ & $17.19 \pm 0.01$ \\ 
58885.5577 & 2.89 & Swift+UVOT & $UVM2$ & $20.30 \pm 0.01$ \\ 
58886.3926 & 3.72 & P48+ZTF & $i$ & $17.52 \pm 0.04$ \\ 
58886.4112 & 3.74 & P48+ZTF & $i$ & $17.52 \pm 0.03$ \\ 
58886.4337 & 3.76 & P60+SEDM & $r$ & $17.20 \pm 0.01$ \\ 
58886.4354 & 3.77 & P60+SEDM & $g$ & $17.34 \pm 0.02$ \\ 
58886.437 & 3.77 & P60+SEDM & $i$ & $17.47 \pm 0.01$ \\ 
58886.4768 & 3.81 & P48+ZTF & $g$ & $17.29 \pm 0.04$ \\ 
58886.4809 & 3.81 & Swift+UVOT & $UVW1$ & $19.17 \pm 0.01$ \\ 
58886.4816 & 3.81 & Swift+UVOT & $U$ & $18.17 \pm 0.01$ \\ 
58886.4819 & 3.81 & Swift+UVOT & $B$ & $17.55 \pm 0.01$ \\ 
58886.4829 & 3.81 & Swift+UVOT & $UVW2$ & $20.05 \pm 0.01$ \\ 
58886.4839 & 3.81 & Swift+UVOT & $V$ & $17.55 \pm 0.01$ \\ 
58886.4854 & 3.82 & Swift+UVOT & $UVM2$ & $20.87 \pm 0.01$ \\ 
58886.4941 & 3.82 & P48+ZTF & $g$ & $17.29 \pm 0.05$ \\ 
58886.5229 & 3.85 & P48+ZTF & $r$ & $17.29 \pm 0.05$ \\ 
58886.5506 & 3.88 & P48+ZTF & $r$ & $17.33 \pm 0.04$ \\ 
58887.2802 & 4.61 & Swift+UVOT & $UVW1$ & $19.48 \pm 0.01$ \\ 
58887.2808 & 4.61 & Swift+UVOT & $U$ & $17.94 \pm 0.01$ \\ 
58887.2812 & 4.61 & Swift+UVOT & $B$ & $17.54 \pm 0.01$ \\ 
58887.2821 & 4.61 & Swift+UVOT & $UVW2$ & $20.47 \pm 0.01$ \\ 
58887.2829 & 4.61 & Swift+UVOT & $V$ & $17.10 \pm 0.01$ \\ 
58887.2842 & 4.61 & Swift+UVOT & $UVM2$ & $20.63 \pm 0.01$ \\ 
58887.3208 & 4.65 & P48+ZTF & $i$ & $17.33 \pm 0.05$ \\ 
58887.429 & 4.76 & P48+ZTF & $g$ & $17.07 \pm 0.04$ \\ 
58887.468 & 4.8 & P48+ZTF & $g$ & $17.10 \pm 0.05$ \\ 
58887.4751 & 4.81 & P48+ZTF & $g$ & $17.10 \pm 0.05$ \\ 
58887.5039 & 4.83 & P48+ZTF & $r$ & $17.07 \pm 0.05$ \\ 
58887.5305 & 4.86 & P48+ZTF & $r$ & $17.08 \pm 0.05$ \\ 
58887.5314 & 4.86 & P48+ZTF & $r$ & $17.05 \pm 0.04$ \\ 
58888.3553 & 5.69 & P60+SEDM & $r$ & $16.81 \pm 0.02$ \\ 
58888.357 & 5.69 & P60+SEDM & $g$ & $16.98 \pm 0.03$ \\ 
58888.36 & 5.69 & P48+ZTF & $i$ & $17.16 \pm 0.04$ \\ 
58888.3928 & 5.72 & P48+ZTF & $i$ & $17.14 \pm 0.05$ \\ 
58888.4746 & 5.8 & P48+ZTF & $r$ & $16.88 \pm 0.04$ \\ 
58888.4892 & 5.82 & P48+ZTF & $r$ & $16.87 \pm 0.05$ \\ 
58888.5373 & 5.87 & P48+ZTF & $g$ & $16.92 \pm 0.05$ \\ 
58888.9397 & 6.27 & Swift+UVOT & $UVW1$ & $19.06 \pm 0.01$ \\ 
58888.9404 & 6.27 & Swift+UVOT & $U$ & $18.29 \pm 0.01$ \\ 
58888.9408 & 6.27 & Swift+UVOT & $B$ & $17.09 \pm 0.01$ \\ 
58888.9418 & 6.27 & Swift+UVOT & $UVW2$ & $20.72 \pm 0.01$ \\ 
58888.9428 & 6.27 & Swift+UVOT & $V$ & $17.08 \pm 0.01$ \\ 
58888.9444 & 6.27 & Swift+UVOT & $UVM2$ & $20.37 \pm 0.01$ \\ 
58890.3717 & 7.7 & P48+ZTF & $i$ & $16.93 \pm 0.03$ \\ 
58890.3941 & 7.72 & P48+ZTF & $i$ & $16.94 \pm 0.03$ \\ 
58890.4565 & 7.79 & P48+ZTF & $r$ & $16.65 \pm 0.04$ \\ 
58890.4747 & 7.8 & P48+ZTF & $r$ & $16.62 \pm 0.06$ \\ 
58890.4756 & 7.81 & P48+ZTF & $r$ & $16.62 \pm 0.04$ \\ 
58890.5276 & 7.86 & P48+ZTF & $g$ & $16.74 \pm 0.05$ \\ 
58890.5588 & 7.89 & P48+ZTF & $g$ & $16.75 \pm 0.05$ \\ 
58890.5597 & 7.89 & P48+ZTF & $g$ & $16.75 \pm 0.05$ \\ 
58891.3937 & 8.72 & P48+ZTF & $i$ & $16.84 \pm 0.03$ \\ 
58891.4157 & 8.75 & P48+ZTF & $i$ & $16.88 \pm 0.03$ \\ 
58891.4552 & 8.79 & P48+ZTF & $g$ & $16.71 \pm 0.04$ \\ 
58891.4626 & 8.79 & P48+ZTF & $g$ & $16.70 \pm 0.04$ \\ 
58891.7595 & 9.09 & Swift+UVOT & $UVW1$ & $19.49 \pm 0.01$ \\ 
58891.7608 & 9.09 & Swift+UVOT & $U$ & $18.22 \pm 0.01$ \\ 
58891.7615 & 9.09 & Swift+UVOT & $B$ & $17.02 \pm 0.01$ \\ 
58891.7634 & 9.09 & Swift+UVOT & $UVW2$ & $20.37 \pm 0.01$ \\ 
58891.7654 & 9.1 & Swift+UVOT & $V$ & $16.44 \pm 0.01$ \\ 
58891.7683 & 9.1 & Swift+UVOT & $UVM2$ & $20.77 \pm 0.01$ \\ 
58892.3651 & 9.7 & P48+ZTF & $g$ & $16.68 \pm 0.05$ \\ 
58892.3832 & 9.71 & P48+ZTF & $g$ & $16.69 \pm 0.04$ \\ 
58892.4559 & 9.79 & P48+ZTF & $i$ & $16.83 \pm 0.03$ \\ 
58892.5181 & 9.85 & P48+ZTF & $r$ & $16.46 \pm 0.04$ \\ 
58892.534 & 9.86 & P48+ZTF & $r$ & $16.45 \pm 0.04$ \\ 
58893.3186 & 10.65 & Swift+UVOT & $UVW2$ & $20.06 \pm 0.01$ \\ 
58893.4023 & 10.73 & P48+ZTF & $i$ & $16.80 \pm 0.03$ \\ 
58893.4715 & 10.8 & P48+ZTF & $g$ & $16.67 \pm 0.04$ \\ 
58893.4965 & 10.83 & P48+ZTF & $g$ & $16.67 \pm 0.04$ \\ 
58893.4974 & 10.83 & P48+ZTF & $g$ & $16.67 \pm 0.04$ \\ 
58893.521 & 10.85 & P48+ZTF & $r$ & $16.41 \pm 0.04$ \\ 
58893.53 & 10.86 & Swift+UVOT & $V$ & $16.49 \pm 0.01$ \\ 
58893.5325 & 10.86 & Swift+UVOT & $UVM2$ & $20.90 \pm 0.01$ \\ 
58893.5338 & 10.86 & P48+ZTF & $r$ & $16.43 \pm 0.03$ \\ 
58893.7579 & 11.09 & Swift+UVOT & $UVW1$ & $19.63 \pm 0.01$ \\ 
58893.759 & 11.09 & Swift+UVOT & $U$ & $18.40 \pm 0.01$ \\ 
58893.7595 & 11.09 & Swift+UVOT & $B$ & $17.11 \pm 0.01$ \\ 
58893.7604 & 11.09 & Swift+UVOT & $UVW2$ & $20.43 \pm 0.01$ \\ 
58894.3388 & 11.67 & P48+ZTF & $g$ & $16.68 \pm 0.04$ \\ 
58894.4351 & 11.77 & P48+ZTF & $i$ & $16.75 \pm 0.03$ \\ 
58894.4554 & 11.79 & P48+ZTF & $i$ & $16.74 \pm 0.03$ \\ 
58894.5153 & 11.85 & P48+ZTF & $r$ & $16.38 \pm 0.04$ \\ 
58894.535 & 11.87 & P48+ZTF & $r$ & $16.37 \pm 0.04$ \\ 
58894.5468 & 11.88 & P48+ZTF & $g$ & $16.70 \pm 0.03$ \\ 
58895.137 & 12.47 & Swift+UVOT & $UVW1$ & $19.89 \pm 0.01$ \\ 
58895.1377 & 12.47 & Swift+UVOT & $U$ & $18.63 \pm 0.01$ \\ 
58895.138 & 12.47 & Swift+UVOT & $B$ & $17.35 \pm 0.01$ \\ 
58895.1391 & 12.47 & Swift+UVOT & $UVW2$ & $20.41 \pm 0.01$ \\ 
58895.14 & 12.47 & Swift+UVOT & $V$ & $16.35 \pm 0.01$ \\ 
58895.1417 & 12.47 & Swift+UVOT & $UVM2$ & $20.97 \pm 0.01$ \\ 
58895.4968 & 12.83 & P48+ZTF & $r$ & $16.35 \pm 0.03$ \\ 
58895.4972 & 12.83 & P48+ZTF & $r$ & $16.33 \pm 0.04$ \\ 
58896.3318 & 13.66 & P48+ZTF & $i$ & $16.72 \pm 0.03$ \\ 
58896.3934 & 13.72 & P48+ZTF & $i$ & $16.70 \pm 0.03$ \\ 
58898.1568 & 15.49 & LT+IOO & $r$ & $16.32 \pm 0.02$ \\ 
58898.1576 & 15.49 & LT+IOO & $i$ & $16.76 \pm 0.02$ \\ 
58898.1585 & 15.49 & LT+IOO & $g$ & $16.75 \pm 0.02$ \\ 
58898.1593 & 15.49 & LT+IOO & $u$ & $18.66 \pm 0.04$ \\ 
58898.445 & 15.77 & P48+ZTF & $g$ & $16.92 \pm 0.04$ \\ 
58898.4558 & 15.79 & P48+ZTF & $g$ & $16.90 \pm 0.03$ \\ 
58898.4955 & 15.83 & P48+ZTF & $r$ & $16.38 \pm 0.03$ \\ 
58898.5119 & 15.84 & P48+ZTF & $r$ & $16.39 \pm 0.04$ \\ 
58898.5128 & 15.84 & P48+ZTF & $r$ & $16.35 \pm 0.04$ \\ 
58898.5335 & 15.86 & P48+ZTF & $r$ & $16.36 \pm 0.03$ \\ 
58898.5463 & 15.88 & P48+ZTF & $g$ & $16.93 \pm 0.04$ \\ 
58899.4051 & 16.74 & P48+ZTF & $g$ & $16.92 \pm 0.04$ \\ 
58899.4351 & 16.77 & P48+ZTF & $g$ & $16.94 \pm 0.04$ \\ 
58899.4828 & 16.81 & P48+ZTF & $r$ & $16.34 \pm 0.04$ \\ 
58899.5057 & 16.84 & P48+ZTF & $r$ & $16.36 \pm 0.03$ \\ 
58899.5302 & 16.86 & P48+ZTF & $g$ & $16.95 \pm 0.05$ \\ 
58900.3929 & 17.72 & P48+ZTF & $g$ & $16.98 \pm 0.05$ \\ 
58900.4467 & 17.78 & P48+ZTF & $r$ & $16.26 \pm 0.04$ \\ 
58900.4499 & 17.78 & P60+SEDM & $r$ & $16.38 \pm 0.01$ \\ 
58900.4516 & 17.78 & P60+SEDM & $g$ & $16.98 \pm 0.02$ \\ 
58900.4532 & 17.78 & P60+SEDM & $i$ & $16.64 \pm 0.02$ \\ 
58900.4787 & 17.81 & P48+ZTF & $r$ & $16.35 \pm 0.03$ \\ 
58900.4938 & 17.82 & P48+ZTF & $r$ & $16.34 \pm 0.04$ \\ 
58900.5289 & 17.86 & P48+ZTF & $g$ & $16.98 \pm 0.07$ \\ 
58901.4137 & 18.74 & P48+ZTF & $r$ & $16.38 \pm 0.03$ \\ 
58901.4335 & 18.76 & P48+ZTF & $r$ & $16.39 \pm 0.03$ \\ 
58901.4546 & 18.78 & P48+ZTF & $r$ & $16.37 \pm 0.03$ \\ 
58901.4546 & 18.78 & P48+ZTF & $r$ & $16.37 \pm 0.03$ \\ 
58902.6701 & 20.0 & Swift+UVOT & $UVW1$ & $20.56 \pm 0.01$ \\ 
58902.6715 & 20.0 & Swift+UVOT & $U$ & $19.52 \pm 0.01$ \\ 
58902.6725 & 20.0 & Swift+UVOT & $B$ & $17.82 \pm 0.01$ \\ 
58902.6748 & 20.0 & Swift+UVOT & $UVW2$ & $20.22 \pm 0.01$ \\ 
58902.6772 & 20.01 & Swift+UVOT & $V$ & $16.49 \pm 0.01$ \\ 
58902.6791 & 20.01 & Swift+UVOT & $UVM2$ & $22.23 \pm 0.01$ \\ 
58903.36 & 20.69 & P48+ZTF & $g$ & $17.21 \pm 0.06$ \\ 
58903.412 & 20.74 & P48+ZTF & $r$ & $16.48 \pm 0.03$ \\ 
58903.4217 & 20.75 & P48+ZTF & $r$ & $16.45 \pm 0.03$ \\ 
58903.4571 & 20.79 & P48+ZTF & $r$ & $16.48 \pm 0.03$ \\ 
58903.4605 & 20.79 & P48+ZTF & $r$ & $16.46 \pm 0.05$ \\ 
58903.4953 & 20.83 & P48+ZTF & $r$ & $16.47 \pm 0.03$ \\ 
58903.4962 & 20.83 & P48+ZTF & $r$ & $16.51 \pm 0.04$ \\ 
58903.5079 & 20.84 & P48+ZTF & $r$ & $16.45 \pm 0.04$ \\ 
58903.5409 & 20.87 & P48+ZTF & $g$ & $17.22 \pm 0.05$ \\ 
58904.3954 & 21.73 & P48+ZTF & $i$ & $16.79 \pm 0.03$ \\ 
58904.4029 & 21.73 & P48+ZTF & $i$ & $16.81 \pm 0.02$ \\ 
58904.4461 & 21.78 & P48+ZTF & $g$ & $17.28 \pm 0.05$ \\ 
58904.489 & 21.82 & P48+ZTF & $r$ & $16.50 \pm 0.03$ \\ 
58906.3392 & 23.67 & P48+ZTF & $g$ & $17.44 \pm 0.05$ \\ 
58906.4339 & 23.76 & P48+ZTF & $i$ & $16.89 \pm 0.02$ \\ 
58906.4868 & 23.82 & P48+ZTF & $r$ & $16.56 \pm 0.04$ \\ 
58906.4878 & 23.82 & P48+ZTF & $r$ & $16.57 \pm 0.03$ \\ 
58906.5057 & 23.84 & P48+ZTF & $r$ & $16.57 \pm 0.03$ \\ 
58906.5381 & 23.87 & P48+ZTF & $g$ & $17.41 \pm 0.05$ \\ 
58906.539 & 23.87 & P48+ZTF & $g$ & $17.45 \pm 0.06$ \\ 
58906.5551 & 23.89 & P48+ZTF & $i$ & $16.88 \pm 0.04$ \\ 
58908.3226 & 25.65 & Swift+UVOT & $UVW1$ & $19.63 \pm 0.01$ \\ 
58908.3236 & 25.65 & Swift+UVOT & $U$ & $20.53 \pm 0.01$ \\ 
58908.3243 & 25.65 & Swift+UVOT & $B$ & $18.19 \pm 0.01$ \\ 
58908.3259 & 25.66 & Swift+UVOT & $UVW2$ & $20.45 \pm 0.01$ \\ 
58908.3275 & 25.66 & Swift+UVOT & $V$ & $16.99 \pm 0.01$ \\ 
58908.3288 & 25.66 & Swift+UVOT & $UVM2$ & $21.54 \pm 0.01$ \\ 
58908.4122 & 25.74 & P48+ZTF & $i$ & $16.98 \pm 0.03$ \\ 
58908.4158 & 25.75 & P60+SEDM & $r$ & $16.77 \pm 0.02$ \\ 
58908.4258 & 25.76 & P48+ZTF & $i$ & $16.99 \pm 0.03$ \\ 
58908.4624 & 25.79 & P48+ZTF & $r$ & $16.73 \pm 0.04$ \\ 
58908.4949 & 25.82 & P48+ZTF & $r$ & $16.76 \pm 0.04$ \\ 
58908.5315 & 25.86 & P48+ZTF & $g$ & $17.46 \pm 0.09$ \\ 
58908.5565 & 25.89 & P48+ZTF & $g$ & $17.45 \pm 0.05$ \\ 
58909.175 & 26.5 & LT+IOO & $r$ & $16.75 \pm 0.02$ \\ 
58909.1758 & 26.51 & LT+IOO & $i$ & $17.03 \pm 0.02$ \\ 
58909.1766 & 26.51 & LT+IOO & $g$ & $17.54 \pm 0.02$ \\ 
58909.1775 & 26.51 & LT+IOO & $u$ & $19.98 \pm 0.07$ \\ 
58909.1789 & 26.51 & LT+IOO & $z$ & $16.66 \pm 0.01$ \\ 
58911.2535 & 28.58 & P48+ZTF & $i$ & $17.07 \pm 0.05$ \\ 
58911.3516 & 28.68 & P48+ZTF & $i$ & $17.14 \pm 0.03$ \\ 
58911.4256 & 28.76 & P48+ZTF & $g$ & $17.85 \pm 0.06$ \\ 
58911.4265 & 28.76 & P48+ZTF & $g$ & $17.89 \pm 0.07$ \\ 
58911.4766 & 28.81 & P48+ZTF & $r$ & $16.87 \pm 0.04$ \\ 
58911.4826 & 28.81 & P48+ZTF & $r$ & $16.87 \pm 0.04$ \\ 
58911.4836 & 28.81 & P48+ZTF & $r$ & $16.85 \pm 0.04$ \\ 
58911.551 & 28.88 & P48+ZTF & $g$ & $17.85 \pm 0.07$ \\ 
58911.5515 & 28.88 & P48+ZTF & $g$ & $17.81 \pm 0.08$ \\ 
58911.5533 & 28.88 & P48+ZTF & $g$ & $17.73 \pm 0.07$ \\ 
58911.5538 & 28.88 & P48+ZTF & $g$ & $17.79 \pm 0.06$ \\ 
58912.1515 & 29.48 & LT+IOO & $r$ & $16.94 \pm 0.02$ \\ 
58912.1523 & 29.48 & LT+IOO & $i$ & $17.21 \pm 0.02$ \\ 
58912.1532 & 29.48 & LT+IOO & $g$ & $17.75 \pm 0.01$ \\ 
58912.154 & 29.48 & LT+IOO & $u$ & $20.15 \pm 0.10$ \\ 
58912.1554 & 29.49 & LT+IOO & $z$ & $16.81 \pm 0.02$ \\ 
58912.3746 & 29.7 & P48+ZTF & $i$ & $17.19 \pm 0.03$ \\ 
58912.3792 & 29.71 & P48+ZTF & $i$ & $17.16 \pm 0.04$ \\ 
58912.4747 & 29.8 & P48+ZTF & $r$ & $16.95 \pm 0.04$ \\ 
58912.4973 & 29.83 & P48+ZTF & $r$ & $16.96 \pm 0.04$ \\ 
58912.5209 & 29.85 & P48+ZTF & $g$ & $17.93 \pm 0.08$ \\ 
58912.5468 & 29.88 & P48+ZTF & $g$ & $17.94 \pm 0.07$ \\ 
\enddata 
\end{deluxetable*} 

\section{Details: mass and radius of the extended material}
\label{sec:appendix-shock-cooling}

This calculation closely follows that of \citet{Kasen2017} and \citet{Nakar2014}.

Assume that the layer undergoing shock cooling has mass $M_e$ and radius $R_e$. Photons diffuse from this layer on a timescale $t_\mathrm{diff}\sim\tau R_e/c$. The layer itself is moving at a characteristic velocity $v_e$: the timescale of expanding is $t_\mathrm{exp}\sim R_e/v_e$.The bulk of photons emerge from the layer where $\tau R_e/c \sim R_e/c$, or $\tau \sim c/v_e$.

At a given radius, the optical depth $\tau$ drops due to expansion: $\tau \sim \kappa \rho R$ where $\rho \sim M_e/(4\pi R^3 / 3)$. The radius increases as $R\sim v_e t$, so we find that $\tau \sim 3 \kappa M_e / (4 \pi (v_e t)^2)$. Setting this equal to $c/v_e$,

\begin{equation}
    t \sim \left( \frac{3}{4 \pi } \frac{\kappa M_e}{v_e c} \right)^{1/2}.
\end{equation}

For SN2020bvc, we have an upper limit on the time to peak of $t_p \lesssim 1\,\days$.
From the spectra, we estimate $v_e \sim 0.1c$.
We take $\kappa = 0.2\,\pcmsq\,\pgram$ for a hydrogen-poor gas.
Altogether, we find $M_e \sim 10^{-2}\,\msol$.
Note that this is an upper limit, because the rise time was likely much faster than what we could measure. So, we conclude that $M_e < 10^{-2}\,\msol$.

Next we estimate $R_e$. We assume that the shock deposits energy $E_\mathrm{dep}$in to the layer.
Then the layer cools from expansion, $E_\mathrm{cool}\sim E_\mathrm{dep} (R_e / v_e t)$. The luminosity from cooling is $L_\mathrm{cool} \sim E_\mathrm{cool}/t_\mathrm{cool} \sim E_\mathrm{dep} R_0/v_e t^2$.

Assuming that the deposited energy is half the kinetic energy $E_\mathrm{KE}$ of the shock, $E_\mathrm{dep}\sim E_\mathrm{KE}/2 = \pi R_e^2 dR \rho v_s^2$, where $dR$ and $\rho$ are the width and density of the layer. Taking $dR \approx R_e$ and $\rho \sim M_e / (4 \pi R_e^2 dR)$ we find $E_\mathrm{dep} \sim v_e^2 M_e/4$. So, our expression for the luminosity is

\begin{equation}
    L_\mathrm{cool} \sim \frac{v_e R_e M_e}{4 t^2}.
\end{equation}

Taking $M_e < 10^{-2}\,\msol$, $t < 1\,\days$, $v_e=0.1c$, and $L > 10^{43}\,\erg\,\psec$,
we find $R_e > 10^{12}\,\cm$.
We can only measure a lower limit on the radius because the true peak luminosity is likely much higher than what we can measure.

\section{Details: properties of the forward shock}
\label{sec:appendix-radio-modeling}

The framework described in \citet{Chevalier1998} assumes that the radio emission arises from a population of relativistic electrons with Lorentz factors that follow a power law of index $p$ down to a cutoff $\gamma_m$,

\begin{equation}
    \frac{dN(\gamma_e)}{d\gamma_e} \propto \gamma_e^{-p}, \gamma \geq \gamma_m,
\end{equation}

\noindent where $2.3 \lesssim p \lesssim 3$ \citep{Jones1991,Pelletier2017}. The expression for the typical electron Lorentz factor $\gamma_m$ is

\begin{equation}
    \gamma_m - 1 \approx \epsilon_e \frac{m_p v^2}{m_e c^2}
\end{equation}

\noindent where $\epsilon_e$ is the fraction of energy in relativitic electrons, $m_p$ is the proton mass, $v$ is the shock velocity, $m_e$ is the electron mass, and $c$ is the speed of light.

The resulting spectrum is a broken power law where $\nu^{5/2}$ at $\nu<\nu_a$ and $\nu^{-(p-1)/2}$ at $\nu > \nu_a$, and $\nu_a$ is called the self-absorption frequency \citep{RL}.
By observing the peak frequency $\nu_p$ and peak flux $F_p$ and assuming that $\nu_p=\nu_a$, we can estimate the outer shock radius $R_p$ and magnetic field strength $B_p$.
We take $p=3$ (the results do not depend strongly on the value of $p$), a filling factor $f=0.5$, and assume equipartition ($\alpha=\epsilon_e/\epsilon_B=1$, where $\epsilon_e/\epsilon_B$ is the ratio of the energy density in relativistic electrons to the energy density in magnetic fields).

Assuming that the radio emission is dominated by the transient, we have an upper limit on the peak frequency of $\nu_p < 3\,\ghz$ and a lower limit on the peak flux density of $F_p > 113\,\mu$Jy at $\Delta t=24\,\days$.
We use Equations (13) and (14) of \citet{Chevalier1998} (C98) to solve for $R$ and $B$,
and find $R > 1.7 \times 10^{16}\,\cm$, $B < 0.34\,\gauss$, and a mean shock velocity up to $13\,\days$ of $v>0.3c$.
Expressions for the total energy thermalized by the shock $U$ and the ambient density $n_e$ are given in \citet{Ho2019cow} (H19), following the same framework as in C98.
Using Equation (12) in H19 and taking $\epsilon_B=1/3$ we find $U=1.3\times10^{47}\,\erg$.
Using Equation (16) in H19 we find $n_e\approx 160 \,\pcmcub$,
which corresponds to a mass-loss rate (Equation (23) of H19) of

\begin{equation}
    \frac{\dot{M}}{v_w} \left(\frac{1000\,\km\,\psec}{10^{-4}\,M_\odot\,\pyr}\right)
    = 0.2
\end{equation}

\noindent where $v_w$ is the wind velocity.

The cooling frequency is defined as

\begin{equation}
    \nu_c = \gamma_c^2 \nu_g,
    \label{eq:nuc}
\end{equation}

where

\begin{equation}
    \gamma_c = \frac{6 \pi m_e c}{\sigma_T B^2 t}
    \label{eq:gammac}
\end{equation}

and 

\begin{equation}
\label{eq:nug}
    \nu_g = \frac{q_e B}{2 \pi m_e c}.
\end{equation}

Combining Equations \ref{eq:nuc}, \ref{eq:gammac}, and \ref{eq:nug}, we have

\begin{equation}
    \nu_c = \frac{18 \pi m_e c q_e}{\sigma_T^2 B^3 t^2} \approx 1.0 \times 10^{13}\,\ghz.
\end{equation}

Finally, we find that the bulk of the electrons have Lorentz factor $\gamma_m = 22$.

\section{Inverse Compton Scattering}
\label{sec:appendix-ic}

The luminosity from inverse Compton scattering of optical photons from the relativistic electrons is

\begin{equation}
    \frac{L_\mathrm{IC}}{L_\mathrm{syn}}=\frac{u_{ph}}{u_B}
\end{equation}

\noindent where $u_\mathrm{ph}$ is the photon energy density (which we measure from our UVOIR observations) and $u_B$ is the magnetic energy density (which we measure from our radio observations; \citealt{RL}).
Taking $R_\mathrm{ph}=2 \times 10^{14}\,\cm$ and $L_\mathrm{bol} > 2 \times 10^{42}\,\erg\,\psec$ we have

\begin{equation}
    u_{ph} = \frac{L_\mathrm{bol}}{4 \pi R^3 /3}  > 0.07\,\erg\,\pcmcub.
\end{equation}

Using $B<0.34\,$G we have 

\begin{equation}
    u_B = \frac{B^2}{8\pi} < 0.005 \,\erg\,\pcmcub
\end{equation}

So, the dominant cooling mechanism is inverse Compton scattering rather than synchrotron radiation, and $L_\mathrm{IC}$ is an order of magnitude greater than $L_\mathrm{syn}$ (the radio luminosity).
Photons emitted at frequency $\nu_0$ that are upscattered by electrons at $\gamma_m$ will emerge with an average frequency $\nu_\mathrm{IC}$ where

\begin{equation}
    \langle \nu_\mathrm{IC} \rangle = \frac{4}{3} \gamma_m^2 \nu_0.
\end{equation}

\vspace{5mm}
\facilities{CXO, Hale, Swift, EVLA, VLA, Liverpool:2m, PO:1.2m, PO:1.5m, NOT}

\software{{\tt CASA} \citep{McMullin2007},
          {\tt astropy} \citep{Astropy2013,Astropy2018},
          {\tt matplotlib} \citep{Hunter2007},
          {\tt scipy} \citep{Virtanen2020},
          {\tt Sherpa} \citep{Freeman2011},
}

\acknowledgements

It is a pleasure to thank the anonymous referee for detailed feedback that greatly improved the clarify and thoroughness of the paper.

A.Y.Q.H. was supported
by the GROWTH project funded by the National Science Foundation under PIRE Grant No.\,1545949,
as well as by the Heising-Simons Foundation.
She would like to thank A. Jaodand and M. Brightman for their assistance with the \chandra\ data reduction and
D. Dong for his help with imaging VLA data.
She would also like to thank D. Khatami and D. Kasen for useful discussions regarding shock-cooling emission,
and E. Ofek for his detailed reading of the manuscript.

R.L. is supported by a Marie Sk\l{}odowska-Curie Individual Fellowship within the Horizon 2020 European Union (EU) Framework Programme for Research and Innovation (H2020-MSCA-IF-2017-794467).
AGY’s research is supported by the EU via ERC grant No. 725161, the ISF GW excellence center, an IMOS space infrastructure grant and BSF/Transformative and GIF grants, as well as The Benoziyo Endowment Fund for the Advancement of Science, the Deloro Institute for Advanced Research in Space and Optics, The Veronika A. Rabl Physics Discretionary Fund, Paul and Tina Gardner, Yeda-Sela and the WIS-CIT joint research grant;  AGY is the recipient of the Helen and Martin Kimmel Award for Innovative Investigation.
C. F. gratefully acknowledges support of his research by the Heising-Simons Foundation
(\#2018-0907).

Based on observations obtained with the Samuel Oschin Telescope 48-inch and the 60-inch Telescope at the Palomar Observatory as part of the Zwicky Transient Facility project. ZTF is supported by the National Science Foundation under Grant No. AST-1440341 and a collaboration including Caltech, IPAC, the Weizmann Institute for Science, the Oskar Klein Center at Stockholm University, the University of Maryland, the University of Washington, Deutsches Elektronen-Synchrotron and Humboldt University, Los Alamos National Laboratories, the TANGO Consortium of Taiwan, the University of Wisconsin at Milwaukee, and Lawrence Berkeley National Laboratories. Operations are conducted by COO, IPAC, and UW.
The scientific results reported in this article are based on
observations made by the Chandra X-ray Observatory.
This research has made use of software provided by the Chandra X-ray Center (CXC) in the application packages CIAO and Sherpa.
This work made use of data supplied by the UK Swift Science Data Centre at the University of Leicester.
SED Machine is based upon work supported by the National Science Foundation under Grant No. 1106171.
The Submillimeter Array is a joint project between the Smithsonian Astrophysical Observatory and the Academia Sinica Institute of Astronomy and Astrophysics and is funded by the Smithsonian Institution and the Academia Sinica.
The Liverpool Telescope is operated on the island of La Palma by Liverpool John Moores University in the Spanish Observatorio del Roque de los Muchachos of the Instituto de Astrofisica de Canarias with financial support from the UK Science and Technology Facilities Council.
Based on observations made with the Nordic Optical Telescope, operated by the Nordic Optical Telescope Scientific Association at the Observatorio del Roque de los Muchachos, La Palma, Spain, of the Instituto de Astrofisica de Canarias.

This work made use of the data products generated by the NYU SN group, and 
released under DOI:10.5281/zenodo.58767, 
available at \url{https://github.com/nyusngroup/SESNspectraLib}.

Funding for the Sloan Digital Sky Survey IV has been provided by the Alfred P. Sloan Foundation, the U.S. Department of Energy Office of Science, and the Participating Institutions. SDSS-IV acknowledges
support and resources from the Center for High-Performance Computing at
the University of Utah. The SDSS web site is www.sdss.org.

SDSS-IV is managed by the Astrophysical Research Consortium for the 
Participating Institutions of the SDSS Collaboration including the 
Brazilian Participation Group, the Carnegie Institution for Science, 
Carnegie Mellon University, the Chilean Participation Group, the French Participation Group, Harvard-Smithsonian Center for Astrophysics, 
Instituto de Astrof\'isica de Canarias, The Johns Hopkins University, Kavli Institute for the Physics and Mathematics of the Universe (IPMU) / 
University of Tokyo, the Korean Participation Group, Lawrence Berkeley National Laboratory, 
Leibniz Institut f\"ur Astrophysik Potsdam (AIP),  
Max-Planck-Institut f\"ur Astronomie (MPIA Heidelberg), 
Max-Planck-Institut f\"ur Astrophysik (MPA Garching), 
Max-Planck-Institut f\"ur Extraterrestrische Physik (MPE), 
National Astronomical Observatories of China, New Mexico State University, 
New York University, University of Notre Dame, 
Observat\'ario Nacional / MCTI, The Ohio State University, 
Pennsylvania State University, Shanghai Astronomical Observatory, 
United Kingdom Participation Group,
Universidad Nacional Aut\'onoma de M\'exico, University of Arizona, 
University of Colorado Boulder, University of Oxford, University of Portsmouth, 
University of Utah, University of Virginia, University of Washington, University of Wisconsin, 
Vanderbilt University, and Yale University.

\bibliography{refs}
\bibliographystyle{aasjournal}

\end{document}